\documentclass[a4paper,11pt]{article}

\pdfoutput=1 
\usepackage{jheppub}   
\usepackage{graphicx}
\usepackage{float}
\usepackage{tensor}
\usepackage[toc,page]{appendix}
\usepackage[usenames,dvipsnames]{xcolor}

\usepackage{amsmath}

\newcommand{\be}{\begin{equation}}
\newcommand{\ee}{\end{equation}}
\newcommand{\ba}{\begin{eqnarray}}
\newcommand{\ea}{\end{eqnarray}}
\newcommand{\nn}{\nonumber}

\newcommand{\tq}{\frak{q}}
\newcommand{\thh}{\frak{h}}

\begin{document}

\title{Hairy black holes in cubic quasi-topological gravity}

\author[a,b]{Hannah Dykaar}
\author[a]{Robie A. Hennigar,}
\author[a]{and Robert B. Mann}

\affiliation[a]{Department of Physics and Astronomy, University of Waterloo, \\
Waterloo, Ontario, Canada, N2L 3G1}
\affiliation[b]{Department of Physics, McGill University,
\\ 3600 rue University, Montreal, QC, H3A 2T8, Canada}
\emailAdd{rhenniga@uwaterloo.ca}
\emailAdd{rbmann@uwaterloo.ca}

\abstract{
We construct a class of five dimensional black hole solutions to cubic quasi-topological gravity with conformal scalar hair and study their thermodynamics.  We find these black holes provide the second example of black hole $\lambda$-lines: a line of second order (continuous) phase transitions, akin to the fluid/superfluid transition of $^4$He.  Examples of isolated critical points are found for spherical black holes, marking the first in the literature to date.  We also find various novel and interesting phase structures, including an isolated critical point occurring in conjunction with a double reentrant phase transition. The AdS vacua of the theory are studied, finding ghost-free configurations where the scalar field takes on a non-zero constant value, in notable contrast to the five dimensional Lovelock case.
}


\maketitle


\section{Introduction}

More than forty years after Hawking's discovery of black hole radiation, black hole thermodynamics continues to provide deep insights into the nature of quantum gravity.  This is especially true for black holes which are asymptotic to anti de Sitter space (AdS) where various gauge/gravity dualities relate the thermodynamic properties of black holes to phenomena in strongly coupled quantum field theories.  Perhaps the most famous example of this kind is the conjectured relationship between the Hawking-Page transition~\cite{Hawking:1982dh} and the deconfinement transition in certain gauge theories~\cite{Witten:1998zw}.

In more recent years there has been growing interesting in the subject of \textit{black hole chemisty} where the cosmological constant is considered as a thermodynamic variable~\cite{Henneaux:1985tv, Creighton:1995au, Caldarelli:1999xj}, with the interpretation of pressure in the first law of black hole mechanics~\cite{Kastor:2009wy, KastorEtal:2010}.  It was found that within this framework there is a physical analogy between the charged AdS black hole and the van der Waals fluid, with the analog of the liquid/gas transition being a small/large black hole phase transition~\cite{Kubiznak:2012wp}.  Numerous other results have since been obtained, including \textit{triple points} and \textit{re-entrant phase transitions} for rotating black holes in $d > 4$~\cite{Altamirano:2013uqa, Altamirano:2013ane}.  The chemistry of black holes in higher curvature gravity~\cite{Wei:2012ui, Cai:2013qga, Xu:2013zea, Mo:2014qsa, Wei:2014hba, Mo:2014mba, Zou:2014mha, Belhaj:2014tga, Xu:2014tja, Frassino:2014pha, Dolan:2014vba, Belhaj:2014eha, Sherkatghanad:2014hda, Hendi:2015cka, Hendi:2015oqa, Hennigar:2015esa, Hendi:2015psa, Nie:2015zia, Hendi:2015pda, Hendi:2015soe, Johnson:2015ekr, Hendi:2016njy, Zeng:2016aly, Hennigar:2016gkm} has proven particularly fruitful, including examples of \textit{multiple re-entrant phase transitions}~\cite{Frassino:2014pha}, \textit{isolated critical points}~\cite{Frassino:2014pha,Dolan:2014vba,Hennigar:2015esa, EricksonRobie} and most recently an analog of a superfluid phase transition~\cite{Hennigar:2016xwd}.  This programme goes beyond critical behaviour with studies developing entropy inequalities for AdS black holes~\cite{Cvetic:2010jb, Hennigar:2014cfa}, discussing the notion of  holographic heat engines~\cite{Johnson:2014yja}, and investigating holographic implications~\cite{Karch:2015rpa, Caceres:2015vsa, Dolan:2016jjc, Couch:2016exn}.  For further details, see ref.~\cite{Kubiznak:2016qmn} for a review.

Here we are concerned with black holes in higher curvature theories of gravity where the Einstein-Hilbert action is supplemented by additional curvature invariants constructed from the Riemann tensor and its contractions.  These terms are expected to arise as low energy corrections from a UV complete theory of gravity, e.g. string theory~\cite{Zwiebach:1985uq}.  From the perspective of holography, higher curvature corrections correspond to $1/N_c$ corrections and allow for the study of a broader class of CFTs~\cite{Buchel:2008vz, Hofman:2009ug, Myers:2010jv}.  However, generic higher curvature theories suffer from problems, such as ghosts, which severely limit the number of sensible theories available.  The most famous example of higher curvature gravity which is ghost free on maximally symmetric backgrounds is \textit{Lovelock gravity}~\cite{Lovelock:1971yv} where the dimensionally extended Euler densities are included in the gravitational action, meaning the $k^{th}$ order contribution is nontrivial in $d \ge 2k+1$.  Lovelock gravity is a natural extension of Einstein gravity to higher dimensions and is the unique higher curvature theory maintaining second order field equations for the metric.  Although Lovelock gravity is the most studied example, other ghost-free higher  curvature theories exist.  For example, it has recently been shown that the most general cubic theory of gravity which propagates the same degrees of freedom as Einstein gravity and has the same form in all dimensions includes, in addition to cubic Lovelock gravity, an additional cubic term which has been dubbed \textit{Einsteinian cubic gravity}~\cite{Bueno:2016xff, Bueno:2016ypa}.

The focus of our study will be hairy black holes in \textit{quasi-topological gravity}~\cite{Oliva:2010eb, Myers:2010ru}.  Quasi-topological gravity is a cubic theory of gravity which is non-trivial in five dimensions, has linearized equations matching those of Einstein gravity, and has second order field equations under the restrictions of spherical symmetry.  On general metrics the field equations are fourth order, and the form of the quasi-topological Lagrangian changes depending on the spacetime dimension.  We consider a non-minimally coupled real scalar field using the method developed by Oliva and Ray~\cite{Oliva:2011np}. This recipe allows one to construct conformally invariant couplings of a scalar field to curvature terms. When the scalar field is coupled to the Euler densities, exact hairy black hole solutions in Einstein and Lovelock gravity  have been found where the scalar field is regular everywhere outside of the horizon and the backreaction of the scalar field onto the metric is captured analytically~\cite{Giribet:2014bva, Giribet:2014fla, Galante:2015voa, Chernicoff:2016jsu, Chernicoff:2016uvq}. These results have provided the first examples of black holes with conformal scalar hair in $d > 4$ where no-go results had been reported previously~\cite{nogo_hairy}. The hairy black holes have been found to have incredibly rich thermodynamics including isolated critical points and black hole $\lambda$-lines~\cite{EricksonRobie, Hennigar:2015wxa,Hennigar:2016xwd}.  Due to the similarity of quasi-topological gravity and Lovelock gravity, our interest here is to see if the interesting results from the Lovelock case can be found in quasi-topological gravity, but in fewer dimensions, or if entirely new results emerge.

Our paper is organized as follows.  In section~\ref{sec:exact_soln} we construct the five dimensional hairy black hole solutions and discuss their thermodynamics.  In section~\ref{sec:criticality} we discuss the relevant physicality constraints and present a general overview of the critical behaviour found in this theory.  We present novel examples of multiple reentrant phase transitions and interesting phase diagrams.  We dedicate two subsections to a discussion of superfluid black hole behaviour and isolated critical points.  We find that the five dimensional black holes present a line of second order phase transitions for suitably chosen coupling constants.  This provides the second example of the superfluid black hole phenomenon first presented in~\cite{Hennigar:2016xwd}.  Furthermore, we find that the scalar hair allows for isolated critical points for black holes with spherical horizons.  This is the first such example, with all previous instances of these critical points occurring for hyperbolic black holes.  These critical points are characterized by non-mean field theory critical exponents and do not coincide with a thermodynamic singularity in general (in contrast to the original situation in which they were first observed \cite{Frassino:2014pha,Dolan:2014vba}).
  In section~\ref{sec:qt_coupling} we consider coupling the scalar field to the quasi-topological term in five dimensions, construct   black hole solutions and discuss their properties. We find that the general critical behaviour remains unaffected by this additional coupling.  In section~\ref{sec:linearized_eqns} we consider the vacua of the theory when a coupling to the quasi-topological density is included.  We find that AdS solutions exist where the scalar field takes on a constant value throughout the spacetime.  The theory propagates the same transverse massless graviton as Einstein gravity on these backgrounds, and the couplings can in general be constrained to ensure the graviton is not a ghost.  In an appendix we present the higher dimensional black hole solutions with coupling to the quasi-topological density.  

\textbf{Note}: As we were finalizing preparation of this manuscript ref.~\cite{Chernicoff:2016qrc} appeared in the literature.  In particular, this paper contains a detailed discussion of the five dimensional black hole solutions of this theory, and also discusses the coupling of the scalar field to both the cubic and quartic quasi-topological densities.  As a result, there is some overlap between the results we present in sections~\ref{sec:exact_soln}, \ref{sec:qt_coupling},  and~\ref{sec:linearized_eqns} and the results of~\cite{Chernicoff:2016qrc}. However there are also many differences: notably, we study AdS vacua for general scalar field configurations, and provide a detailed account of the thermodynamics from the perspective of black hole chemistry.

\section{Exact Solution \& Thermodynamics}
\label{sec:exact_soln}

We consider a theory containing a Maxwell field and a real scalar field conformally coupled to gravity through a non-minimal coupling between the scalar field and the dimensionally extended Euler densities.  The theory is conveniently written in terms of the rank four tensor,
\ba 
\tensor{S}{_\mu_\nu^\gamma^\delta} &=& \phi^2 \tensor{R}{_\mu_\nu^\gamma^\delta} - 2 \delta^{[\gamma}_{[\mu} \delta^{\delta]}_{\nu]}\nabla_\rho\phi\nabla^\rho\phi 
- 4 \phi \delta^{[\gamma}_{[\mu} \nabla_{\nu]} \nabla^{\delta]} \phi + 8 \delta^{[\gamma}_{[\mu}\nabla_{\nu]}\phi\nabla^{\delta]}\phi \,.
\ea
By construction, $\tensor{S}{_\mu_\nu^\gamma^\delta}$ has the same symmetries as the Riemann tensor.

Here our goal will be to construct hairy black hole solutions of quasi-topological gravity and study their properties.  To this end, we consider the following action,
\ba 
{\cal I} &=& \frac{1}{16\pi} \int d^5x \sqrt{-g} \bigg[-2\Lambda + R + \frac{\lambda}{2} {\cal L}_2 - \frac{7}{4} \mu{\cal Z}_5 - F^2
\nn\\ && \hspace{80pt}+ 16 \pi \left( b_0 \phi^5 S^{(0)} + b_1 \phi S^{(1)} + b_2 \phi^{-3} S^{(2)} \right)  \bigg]
\ea 
where ${\cal L}_2$ is the Gauss-Bonnet Lagrangian density,
\be 
{\cal L}_2 = R^2 -4 \tensor{R}{_\mu _\nu}\tensor{R}{^\mu ^\nu} + \tensor{R}{_\nu _\mu _\rho _\lambda} \tensor{R}{^\nu ^\mu ^\rho ^\lambda}\, ,
\ee
 and ${\cal Z}_D$ is the cubic quasi-topological Lagrangian density in $D$ dimensions,
\begin{align}
\mathcal{Z}_D &= \left[ \vphantom{\left( \frac{3(3D - 8)}{8} R_{\mu \alpha \nu \beta} R^{\mu \alpha \nu \beta} R \right.} {{{R_\mu}^\nu}_\alpha}^\beta {{{R_\nu}^\tau}_\beta}^\sigma {{{R_\tau}^\mu}_\sigma}^\alpha \right.  
               + \frac{1}{(2D - 3)(D - 4)} \left( \frac{3(3D - 8)}{8} R_{\mu \alpha \nu \beta} R^{\mu \alpha \nu \beta} R \right. \nonumber \\
              &  - 3(D-2) R_{\mu \alpha \nu \beta} {R^{\mu \alpha \nu}}_\tau R^{\beta \tau} + 3D\cdot R_{\mu \alpha \nu \beta} R^{\mu \nu} R^{\alpha \beta} 
               \left. + 6(D-2) {R_\mu}^\alpha {R_\alpha}^\nu {R_\nu}^\mu \right. \nonumber \\
              & \left. \left.- \frac{3(3D-4)}{2} {R_\mu}^\alpha {R_\alpha}^\mu R + \frac{3D}{8} R^3 \right) \right]
\label{quasitop}
\end{align}
which here we have considered for the specific case $D=5$.  The scalar field is coupled to gravity via the following terms,
\ba 
S^{(0)} &=& 1\, ,
\nn\\
S^{(1)} &=& S = \tensor{g}{^\mu ^\nu}\tensor{S}{_\mu _\nu} = \tensor{g}{^\mu ^\nu}\tensor{S}{^\rho _\mu _\rho _\nu}\, ,
\nn\\
S^{(2)} &=& S^2 -4 \tensor{S}{_\mu _\nu}\tensor{S}{^\mu ^\nu} + \tensor{S}{_\nu _\mu _\rho _\lambda} \tensor{S}{^\nu ^\mu ^\rho ^\lambda}\, .
\ea

We consider a static, spherically symmetric metric ansatz of the form,
\be\label{metric_ansatz} 
ds^2 = - \frac{r^2}{l^2} f(r) dt^2 + \frac{l^2 dr^2}{r^2 g(r)} + r^2 d \Omega^2_{3(k)}
\ee
where $d \Omega^2_{3(k)}$ is the line element of a 3-dimensional hypersurface of constant positive, zero or negative curvature ($k=1,0,-1$, respectively). We parametrize the electromagnetic potential as,
\be\label{em_ansatz} 
A = e\frac{r}{l}E(r) dt
\ee
from which the field strength follows in the standard way, $F_{\mu \nu } = \partial_\mu A_\nu - \partial_\nu A_\mu$.  
The scalar field must obey the equation,
\be 
5b_0 \phi^4 + 3b_1 S^{(1)} + b_2 \phi^{-4} S^{(2)} = 0
\ee
which follows from the variation of the action with respect to $\phi$ and ensures that the trace of the energy momentum tensor of the scalar field vanishes on-shell.  

We obtain the field equations in the following manner. First, the ansatzes \eqref{metric_ansatz} and \eqref{em_ansatz} are substituted into the action, followed by integrating by parts several times to remove second order derivatives of $f(r)$ and $g(r)$.  We then make the substitution $f(r) = N^2(r)g(r)$ and integrate by parts again to remove first derivatives of $N(r)$.  The resulting action is then varied with respect to $g(r)$, $N(r)$ and $E(r)$ to obtain the gravitational and electromagnetic equations of motion. Finally, we arrive at the set of field equations,
\begin{align}\label{field-eqs} 
\left(1 - 2\frac{\lambda}{l^2} \kappa + \frac{3 \mu}{l^4} \kappa^2 \right)N'(r) &= 0
\nn\\
\left[3r^4\left(1 - \kappa + \frac{\lambda}{l^2} \kappa^2 - \frac{\mu}{l^4}\kappa^3 \right) \right]' 
&= \frac{2 e^2 r^3}{N^2(r)}\left[(rE)' \right]^2 - \frac{16 \pi l^2 h}{r^2} 
\nn\\
\left( \frac{r^3 (rE)'}{N(r)} \right)' &= 0
\end{align}
where we have used the shorthand $\kappa = g(r) - k l^2/r^2$. 

Meanwhile, to satisfy the equations of motion of the scalar field, the following relationships must hold: 
\begin{align}\label{eqn:scalar_field_constraints}
\hspace{50pt} \phi &= \frac{K}{r}  \, , & & &
K &= \varepsilon \sqrt{\frac{-18 b_1 k}{5b_0}} \, , \hspace{90pt}
\nn\\
h &= K^3\left(b_0K^2 + 6kb_1 \right) \, ,  & & &
9 b_1^2  &=  10b_2b_0 \, , 
\end{align}
where $\varepsilon  = -1,0,1$.  It is easy to see from these relationships that, when $k=0$, the scalar field configuration vanishes.  Therefore, throughout this work we shall focus only on black hole solutions of spherical and hyperbolic horizon topologies. 

In solving the field equations, we shall use $N(r)=1$ to solve the first equation in eq.~\eqref{field-eqs}.  The third equation can then be integrated to give,
\be 
A_t = \frac{\sqrt{3}}{2} \frac{e}{r^2}
\ee
while the remaining relationship yields,
\be\label{master-eqn} 
1 - \tilde{\kappa} + \frac{\lambda}{l^2}\tilde\kappa^2 - \frac{\mu}{l^4}\tilde\kappa^3 = \frac{l^2 m}{r^4} - \frac{e^2l^2}{r^6}  + \frac{16 \pi l^2 h}{3 r^5}
\ee
where
\be 
\tilde \kappa = \frac{l^2}{r^2}\left(F(r) - k \right)
\ee
with $F(r) = r^2g(r)/l^2$ defined so  that the metric reads: $-F(r)dt^2 + \cdots$.  Furthermore, note that we have defined the mass and electric charge parameters, $m$ and $e$, for consistency with \cite{Hennigar:2015esa}.  The effect of the backreaction of the scalar field on the spacetime geometry is captured through the ``hairy parameter" $h$, which can take positive and negative values.

With the solution obtained, we can now consider its thermodynamics; in doing so, we find it most convenient to work directly with eq.~\eqref{master-eqn}.  Working in extended phase space, we employ the extended first law,
\be\label{f-law} 
dM = TdS + VdP + \Phi dQ + \Psi_\lambda d\lambda + \Psi_\mu d\mu + {\cal K} dh \, ,
\ee
where, as in \cite{KastorEtal:2010}, we have considered the variation of all dimensionful quantities. In the first law, the thermodynamic mass, charge and electric potential are given by 
\begin{align}
M = \frac{3}{16 \pi} \omega_{3(k)} m \, , \quad
Q = \frac{\sqrt{3}}{4\pi} \omega_{3(k)} e \, ,
\quad
\Phi = \frac{\sqrt{3}}{2} \frac{e}{r^2} \, .
\end{align}
In the above, $\omega_{3(k)}$ is the volume of $d\Omega_{3(k)}^2$, which in the case $k=1$ is $\omega_{3(1)} = 2\pi^2$.  

We identify the cosmological constant as a thermodynamic pressure,
\be\label{cosmological}
P = - \frac{\Lambda}{8\pi} = \frac{3}{4\pi l^2} \, ,
\ee
and find that consistency of the first law requires
\begin{align}
V &= \frac{\omega_{3(k)}}{4} r_+^4 \, ,
\nn\\
\Psi_\lambda &=\frac{3 \omega_{3(k)}}{16 \pi} k \left(k - 8\pi r_+ T \right) \, , 
\nn\\
\Psi_\mu &= \frac{3 \omega_{3(k)}}{16 \pi} k^2 r_+^{-2} \left( k + 12\pi r_+ T \right) \, ,
\nn\\
{\cal K} &= \frac{1}{3h} \left[2\left(M+PV-\Phi Q - \lambda \Psi_\lambda \right) - 3TS - 4\mu\Psi_\mu \right]
\end{align}
where $T$ represents the black hole temperature. This is most easily obtained using Euclidean methods, Wick rotating the time coordinate and requiring regularity of the analytically continued metric,
\be 
T = \frac{1}{4\pi} \left[\frac{\partial F(r)}{\partial r} \right]_{r=r_+} \, .
\ee  
Differentiating eq.~\eqref{master-eqn} and solving for the derivative yields,
\begin{align}\label{temper}
T &= \frac{1}{4\pi \left(r_+^4 +2k\lambda r_+^2 + 3k^2\mu \right)}\left[4 \frac{r_+^5}{l^2} + 2k r_+^3 + \frac{16\pi}{3} h - 2 \frac{e^2}{r_+} - \frac{2 k^3 \mu}{r_+} \right] \, .
\end{align}
The entropy can be calculated using Wald's prescription \cite{Wald:1993nt} and is given by,
\be\label{entropy} 
S =   \frac{A}{4}\left[1+\frac{6k\lambda}{r_+^2} -\frac{9k^2\mu}{r_+^4}   - \frac{40 \pi k h}{3  r_+^3} \right]\, , \quad A = \omega_{3(k)} r_+^3 \, .
\ee

The thermodynamic quantities  satisfy both the first law eq.~\eqref{f-law} and the extended Smarr relation
\be 
2M = 3TS-2PV+2\Phi Q + 2\lambda\Psi_\lambda + 4 \mu \Psi_\mu + 3h{\cal K} 
\ee
 which is consistent with Eulerian scaling.

\section{Criticality}
\label{sec:criticality}

\subsection{Equations and constraints}

To study critical behaviour it is advantageous to work in terms of dimensionless quantities. To this end, we define
\begin{align}\label{parameters}
r_+ &= v |\mu|^{1/4} \, , \quad 
t = 3|\mu|^{1/4} T \, , \quad 
p = 4 \sqrt{|\mu|} P \, ,\quad 
\tq = \frac{e}{2} \sqrt{\frac{6}{|\mu| \pi}} \, , \quad 
\thh = 4 h |\mu|^{-3/4}\, , \quad
\alpha = \frac{\lambda}{\sqrt{|\mu|}} \, .
\end{align}
In terms of these quantities, the equation of state reads,
\be 
p = \frac{t}{v} - \frac{3k}{2 \pi v^2} + \frac{2 \alpha k t}{v^3} + \epsilon \frac{3 k^2 t}{v^5} + \epsilon \frac{3 k^3}{2 \pi v^6} + \frac{\tq^2}{v^6} - \frac{\thh}{v^5} \, ,
\ee
where $\epsilon = {\rm sign}(\mu)$ represents the sign of the quasi-topological coupling.  We will also consider the Gibbs free energy, $G = M - TS$, which we make dimensionless through the following rescaling,
\be 
g = \frac{\mu^{-1/2}}{\omega_{3(k)}} G \, .
\ee
In defining $g$ we have divided by $\omega_{3(k)}$. This ensures the Gibbs free energy is well defined for $k=-1$, where $\omega_{3(-1)}$ is infinite (when the transverse space is non-compact) or not easily computed.  The dimensionless Gibbs energy is given by
\begin{align}
g &= -\frac{1}{16 \pi \left(v^4 + 2\alpha k v^2 +  3 \epsilon k^2 \right) } \bigg[\frac{9 k^5}{v^2}  -  27 k^4 \epsilon \alpha    -  2k^3(3\alpha^2 +  16\epsilon ) v^2 - 3k^2 (5\pi \epsilon p -  \alpha) v^4  
\nn\\ 
 &+  k(6\pi \alpha p - 1) v^6  + \frac{\pi}{3} p v^8 \bigg] + \frac{\tq^2}{4 \left( v^4 + 2\alpha k v^2 + 3 \epsilon k^2 \right)} \left[-\epsilon\frac{3 k^2}{2 v^2} + 3k\alpha + \frac{5 v^2}{6}  - \frac{10 \pi k \thh}{9 v }\right]
 \nn\\
 &+ \frac{k \thh}{36  \left( v^4 + 2\alpha k v^2 + 3 \epsilon k^2\right)} \left[-\epsilon\frac{15 k^3}{v} + 10 \pi \thh  -36  \alpha v +  15 v^3 k - \frac{12 v^3}{k} + 10 \pi p v^5 \right]
\end{align}

It is clear from eq.~\eqref{entropy} that for some parameter values the entropy of the black holes can be negative.  In terms of our dimensionless variables, the requirement for having positive entropy is given by the following inequality:
\be 
1 + \frac{6k\alpha}{v^2} - \epsilon \frac{9 k^2}{v^4} - \frac{10 \pi k \thh}{3 v^3} > 0 \, . 
\ee
If one takes the stance that the Iyer-Wald entropy is counting the true microscopic degrees of freedom of the black hole, then this inequality should be elevated to the level of a physicality constraint---that is, one should exclude the negative entropy solutions from the study.  However, often it is the case that there is no other pathological behaviour associated with the negative entropy solutions, and overall the physical status of these solutions remains unclear.  In what follows we will employ the positivity of entropy condition.

We must also ensure that the black holes under study possess asymptotic regions---something which is not generally true for black holes in higher curvature gravity.  To ensure proper asymptotics we study the discriminant of the field equation~\eqref{master-eqn}, considered as a function of $F(r)$, at large $r$.  Expressing the result in terms of dimensionless parameters  we find
\be\label{eqn:asymptotic_discriminant} 
\frac{\Delta}{|\mu|^3 \!} = -\frac{243}{\pi^4 p^4} \left[\pi^2 p^2 + \frac{4\pi}{9} \left(\alpha^2 - \frac{9}{2} \epsilon \right) \alpha p - \frac{\alpha^2}{3} + \frac{4 \epsilon}{3} \right] \, ,
\ee
which vanishes at the pressures
\be 
p_\pm = \frac{1}{9 \pi } \left[-2\alpha^3 + 9  \alpha \epsilon \pm 2 \left(\alpha^2 - 3\epsilon \right)^{3/2} \right] \, .
\ee
From the theory of polynomials we know that for $\Delta \ge 0$ there are three real roots, with (at least) two of these coinciding when $\Delta = 0$; for $\Delta < 0$, there is only a single real root. By restricting to the cases where all branches admit asymptotics (i.e. where $\Delta \ge 0$), we ensure that the solutions studied are well-behaved.  

The conditions under which $\Delta > 0$ depends on the value of $\epsilon$. For $\epsilon=+1$, corresponding to positive quasi-topological coupling, the positivity of the discriminant can be expressed in the following form:
\be 
\Delta > 0 \Rightarrow \begin{cases} 
p \in (0, p_+) & \text{for } \alpha \in (-\infty, -2) \,,
\\
p \in \left({\rm max} \left\{0, p_- \right\}, p_+ \right) & \text{for } \alpha \in (\sqrt{3}, \infty) \, . 
\end{cases}
\ee  
Note that this indicates that for $\alpha \in (-2, \sqrt{3})$ the discriminant is negative and there is only a single real root: this corresponds to the quasi-topological branch, which does not admit smooth $\mu \to 0$ and $\lambda \to 0$ limits. In the case $\epsilon = -1$, the constraint for asymptotics is simply stated as $p \in (0, p_+)$. 

Ensuring asymptotics is not sufficient: we must also ensure that, for a given choice of coupling constants, the theory is asymptotic to AdS and free from ghost instabilities on this maximally symmetric background.  We defer a complete discussion of this subject to section~\ref{sec:linearized_eqns}, simply quoting the relevant results here.  The simplest and perhaps most natural case to consider is that where the equations of motion are expanded about a background in which the scalar field vanishes.  In this case, the theory is free of ghosts provided that,
\be 
1 - 2 \frac{\lambda}{l^2} F_{\infty} + 3 \frac{\mu}{l^4} F^2_{\infty} > 0 \, ,
\ee
where $l/\sqrt{F_\infty}$ denotes the length scale of the AdS background with $F_\infty$ being the asymptotic value of the metric function.  In terms of the dimensionless quantities we are employing, this inequality becomes
\be\label{eqn:ghost_ineq} 
1 - \frac{2}{3} \pi \alpha p F_{\infty} + \epsilon \frac{\pi^2 \!}{3} p^2 F_{\infty}^2 > 0\, .
\ee
On a maximally symmetric background $F_{\infty}$ is determined by the polynomial equation,
\be 
1 - F_\infty + \frac{\lambda}{l^2} F_\infty^2 - \frac{\mu}{l^4} F_\infty^3 = 0 \, ,
\ee
which corresponds to AdS when $F_\infty > 0$.  Due to this, the results for a ghost free AdS vacuum reduces to the same conditions discussed in~\cite{Hennigar:2015esa}, meaning there are black hole solutions in the regions where $\Delta > 0$.

\subsection{$P-v$ criticality}

Having now discussed the physicality constraints, we move on to a survey of the critical behaviour for these black holes.  We divide this discussion into two short subsections which contain general results for the critical behaviour in the cases of positive and negative coupling.  In these subsections we do not attempt to be exhaustive, but rather highlight some of the salient and interesting features of these black holes.  We also include in two separate sections discussions of superfluid black holes and isolated critical points, which are the most interesting results of this study.  

\subsubsection{Positive $\mu$ thermodynamics}

\begin{figure}[htp]
\centering
\includegraphics[width=0.4\textwidth]{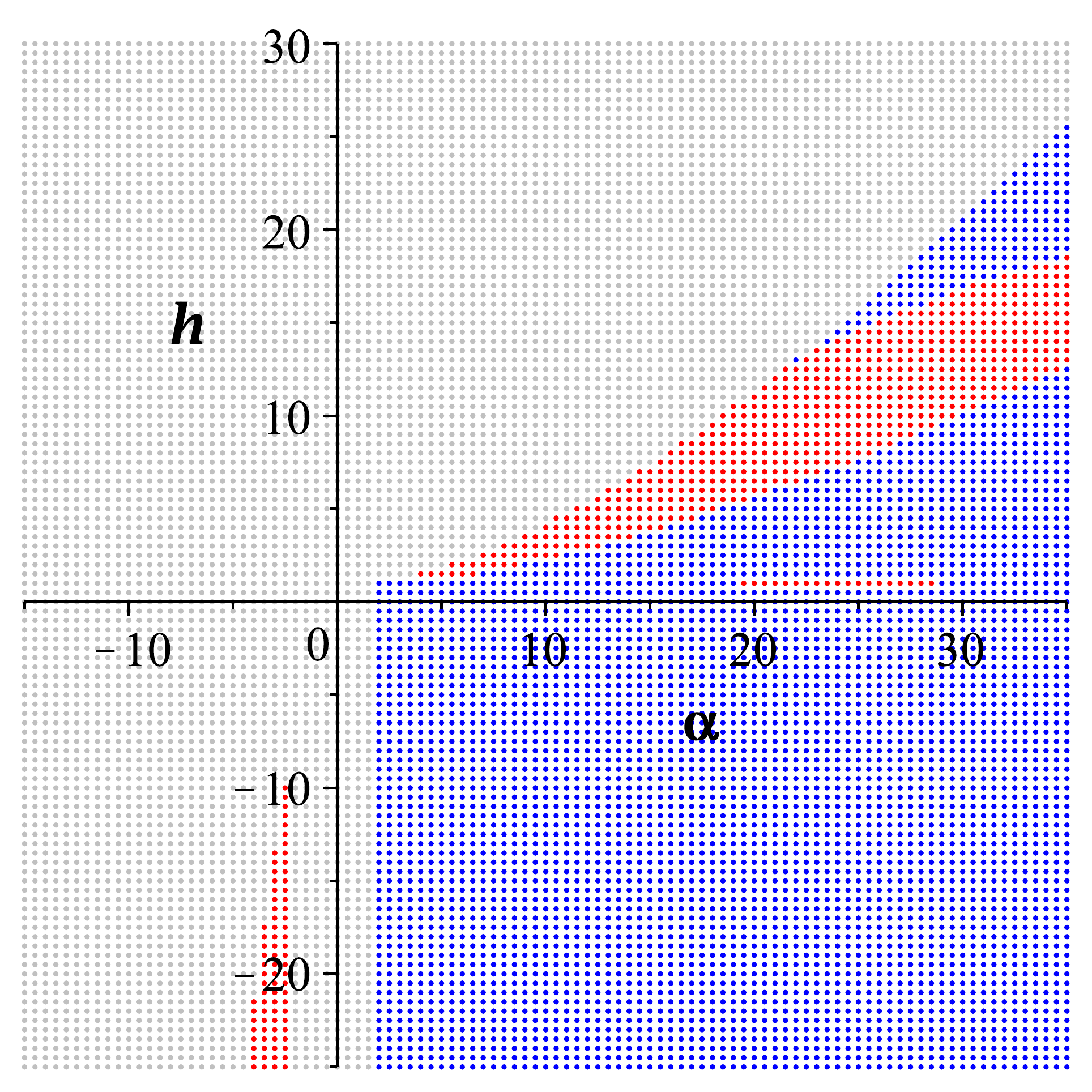}
\includegraphics[width=0.4\textwidth]{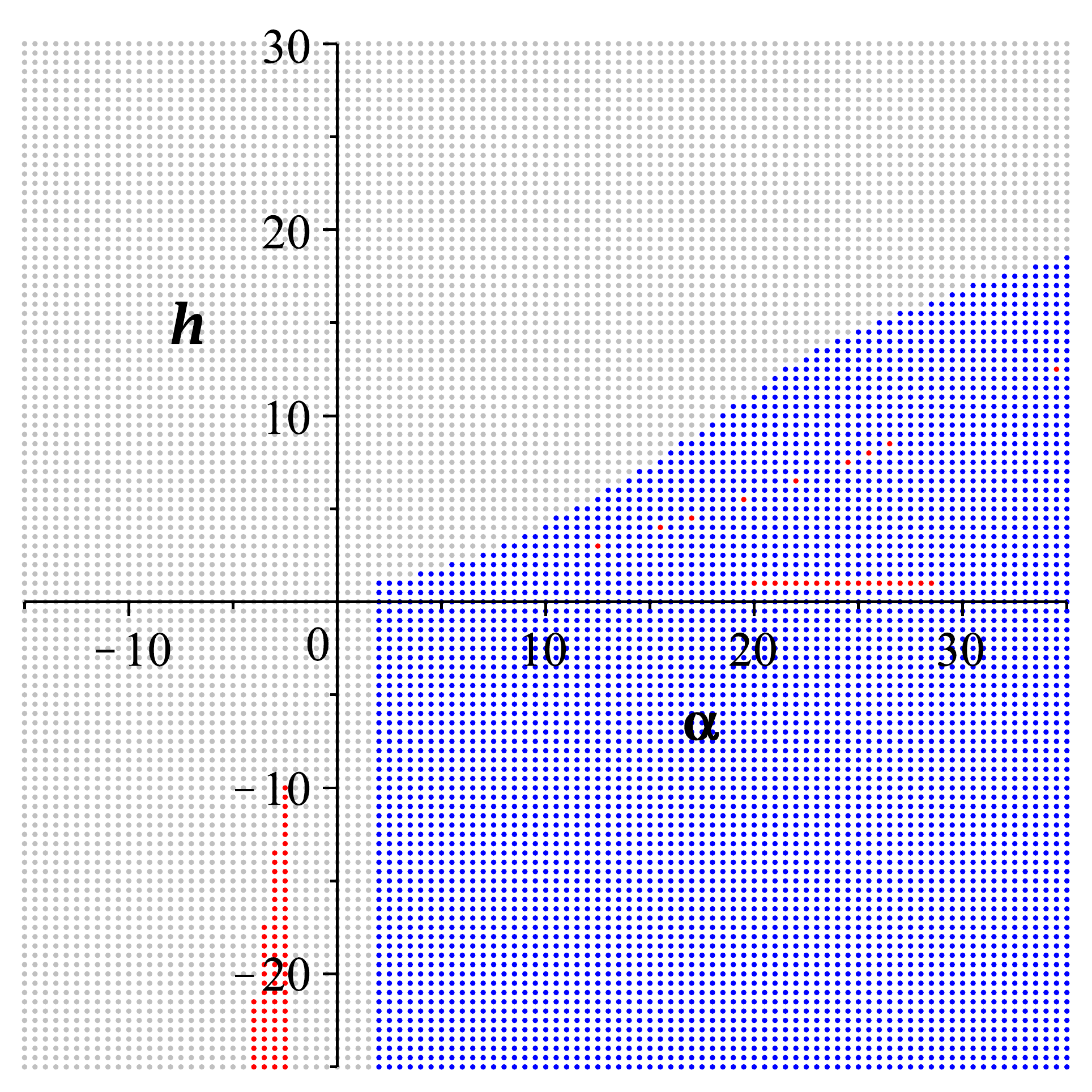}
\caption{{\bf Possible critical points in $(\alpha,\thh)$ parameter space for $k = 1$, $\epsilon = +1$ and $q = 0$}: Here red, blue, and grey points indicate two, one, and zero (possible) critical points, respectively.  \textit{Left}: Possible critical points satisfying entropy and pressure conditions. \textit{Right}: Actual critical points, taking into account only those which occur on the minimizing branch of the Gibbs free energy: we see a vast reduction of the parameter space corresponding to two critical points.  }
\label{fig:pos_mu_number_critical_points_k=1}
\end{figure}

\begin{figure}[htp]
\centering
\includegraphics[width=0.3\textwidth]{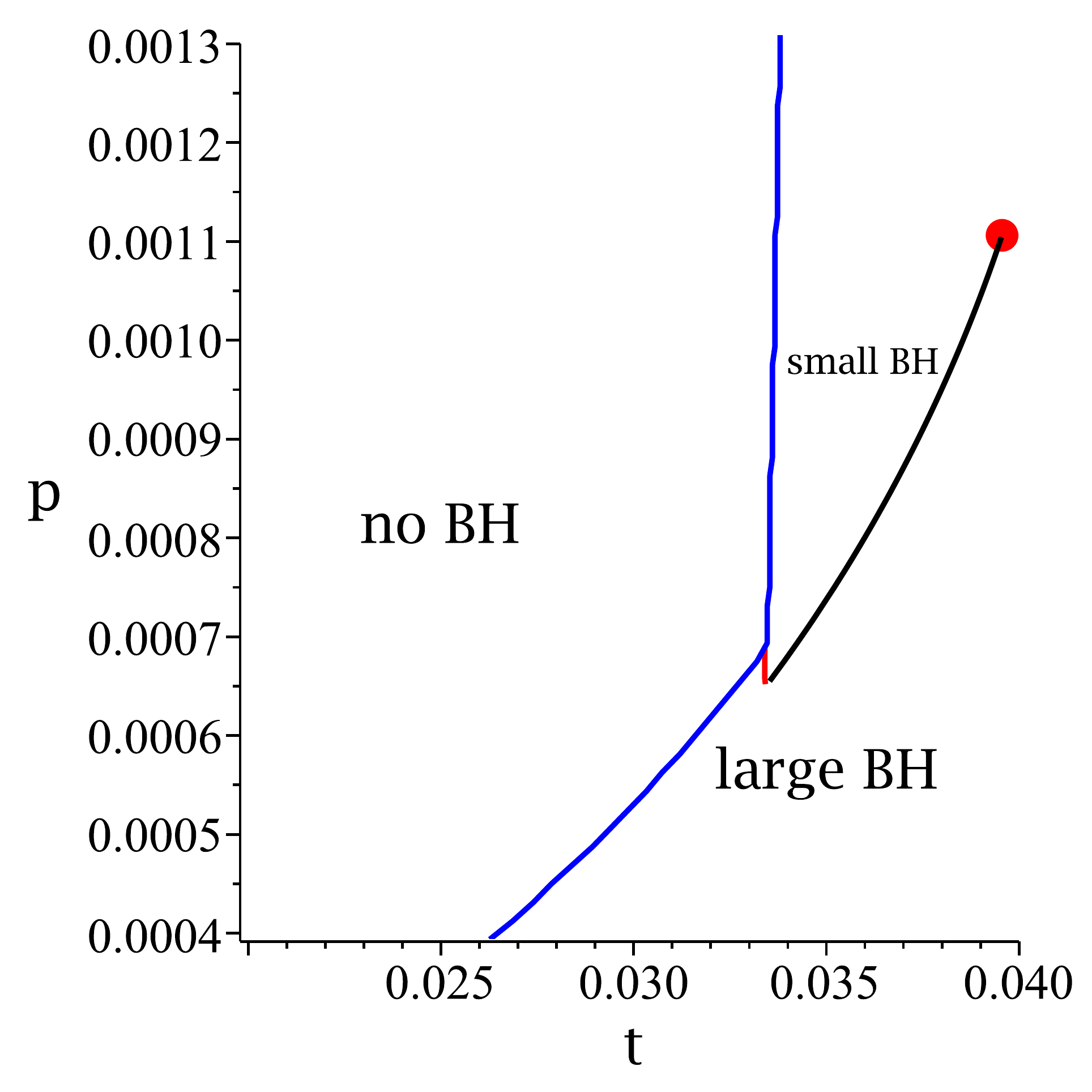}
\includegraphics[width=0.3\textwidth]{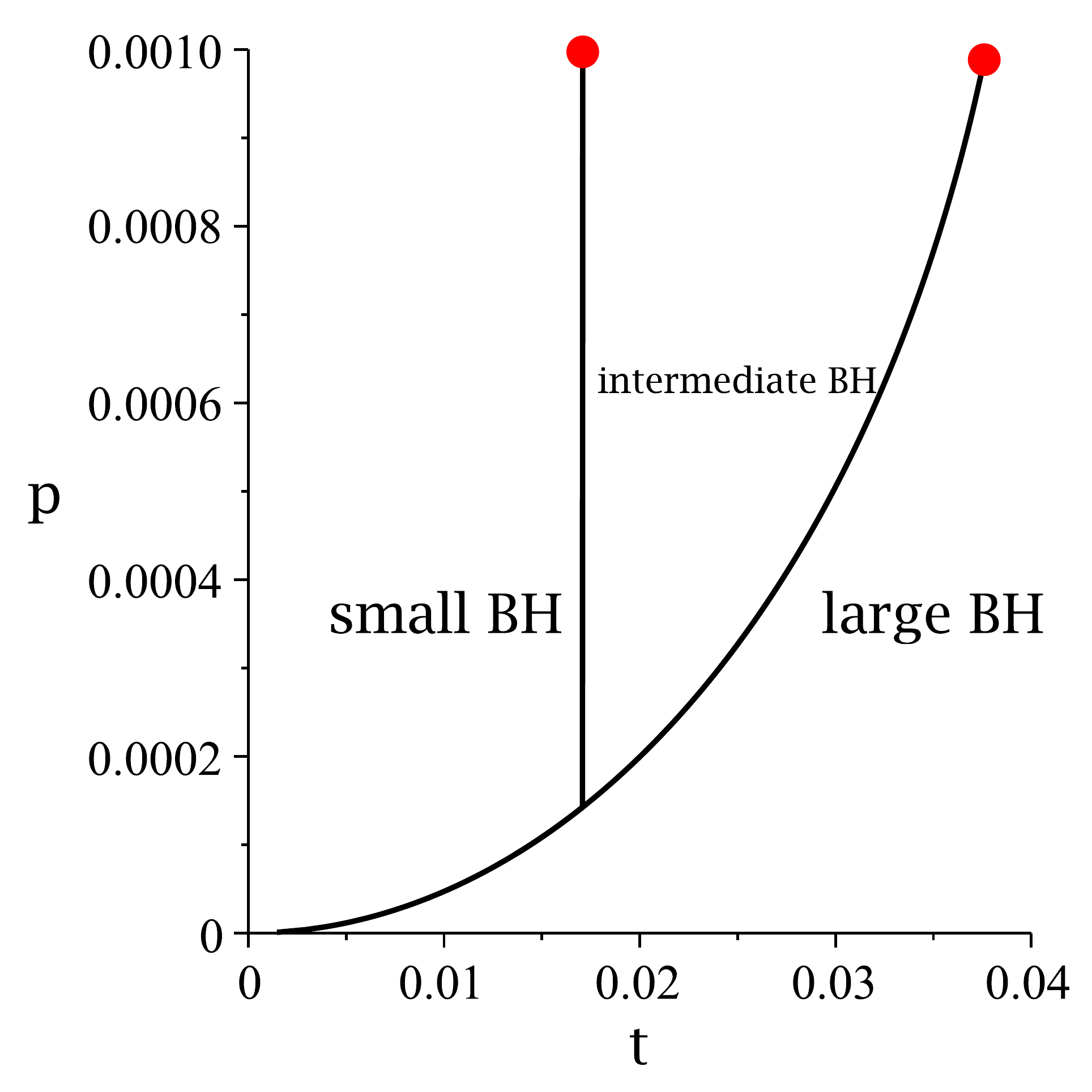}
\includegraphics[width=0.3\textwidth]{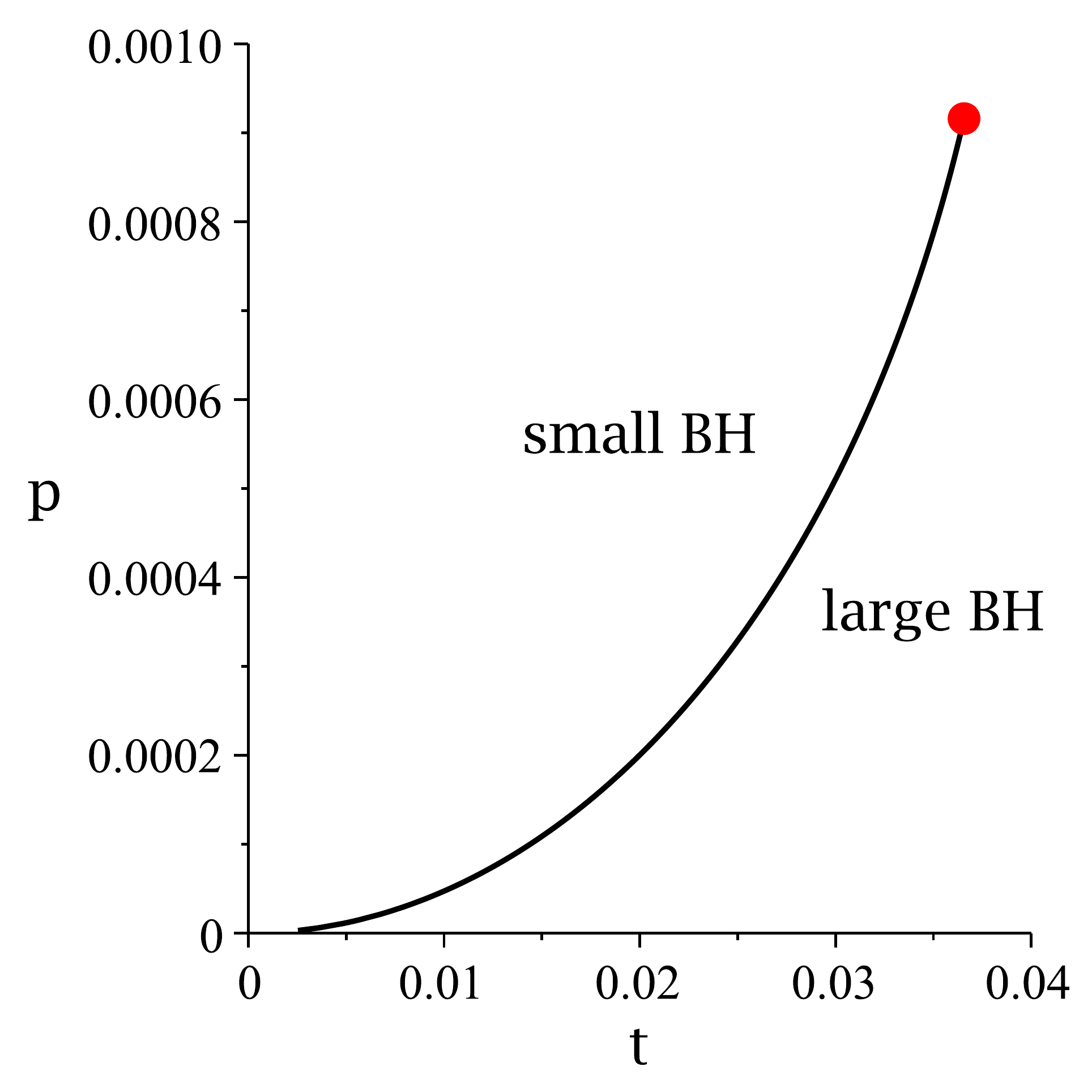}
\caption{{\bf Phase plots for $k = 1$, $\epsilon = +1$ and $\tq = 0$}: \textit{Left}: plot for $\alpha = 25$ and $\thh = 5$. The black curve denotes the coexistence line which ends at the physical critical point (red point). The blue curve denotes the boundary of the ``no black hole region" and the (small) red line denotes a zeroth-order phase transition. For pressures in this region, a large-small-large reentrant phase transition takes place. \textit{Center}: plot for $\alpha = 27$ and $\thh = 0.9899$ showing a triple point where the three coexistence curves intersect. \textit{Right}: plot for $\alpha = 27$ and $\thh =- 11$ showing a single physical critical point and coexistence curve displaying van der Waals type behaviour.}
\label{fig:pos_mu_coexistence_curves_k=1}
\end{figure}

Considering first the case of positive quasi-topological coupling ($\epsilon = +1$) vanishing electric charge and spherical horizons ($k = +1)$, we find that the space of possible critical points takes the form shown in the left plot of figure~\ref{fig:pos_mu_number_critical_points_k=1}. This plot has been constructed by searching for parameters where the equation of state satisfies the conditions for a critical point,
\be 
\frac{\partial p}{\partial v} = 0 = \frac{\partial^2 p}{\partial v^2} \, . 
\ee
However, just having candidate parameters which satisfy these conditions does not guarantee the candidate is a bonafide critical point.  Determining this requires a study of the Gibbs free energy of the black holes, ensuring that the specified thermodynamic parameters correspond to a local minimum of the Gibbs free energy---this is the reason for referring to these as ``possible" critical points.   In the same figure we show what this parameter space looks like when the requirement of minimizing the Gibbs free energy is taken into account, which results in a significant reduction of the available parameter space where two critical points occur. Exploring each of these regions we find a variety of interesting thermodynamic behaviour, including van der Waals behaviour, re-entrant phase transitions and triple points, as we display in figure~\ref{fig:pos_mu_coexistence_curves_k=1}.  

In fact, the narrow range of two critical points in figure~\ref{fig:pos_mu_number_critical_points_k=1} given by the red `bar' in the upper-right quadrant [approximately, $\alpha \in (20, 30)$] corresponds to a region where we see the ``virtual triple point" (VTP) behaviour first pointed out in~\cite{EricksonRobie}.  More precisely, starting in the blue region with $\thh < 0.9884$ we see van der Waals behaviour. As $\thh$ is increased and enters the interval $\thh \in (0.9884, 1.0028)$ we see a triple point akin to that illustrated in figure~\ref{fig:pos_mu_coexistence_curves_k=1} emerge and then disappear, to leave again van der Waals behaviour for $\thh > 1.0028$.  At each boundary of the region of two critical points (i.e. $\thh = 0.9884$ and $ 1.0028$) there is a coexistence curve which displays van der Waals type behaviour, but with a second critical point existing precisely on the coexistence curve (cf. figure~(10) in~\cite{EricksonRobie}).  One can imagine this as the one of the `branches' of the coexistence curves in the triple point diagram shown in the center plot of figure~\ref{fig:pos_mu_coexistence_curves_k=1} shrinking and ultimately the critical point touching the other coexistence curve, which corresponds to the VTP.  More concisely, we can write that the following sequence of critical behavour occurs
\[ 
\text{VdW} \to \text{VTP} \to \text{TP} \to \text{VTP} 
\to \text{VdW} 
\]
as $\thh$ is moved into this interval and then through it.   

In the case of hyperbolic black holes ($k=-1$) with vanishing electric charge the situation appears quite simple: any critical points that occur violate at least one of the physicality conditions.  That is, a plot of the kind shown in figure~\ref{fig:pos_mu_number_critical_points_k=1} would be entirely filled with grey points when the positivity of entropy and pressure constraints are enforced.

\begin{figure}[htp]
\centering
\includegraphics[width=0.3\textwidth]{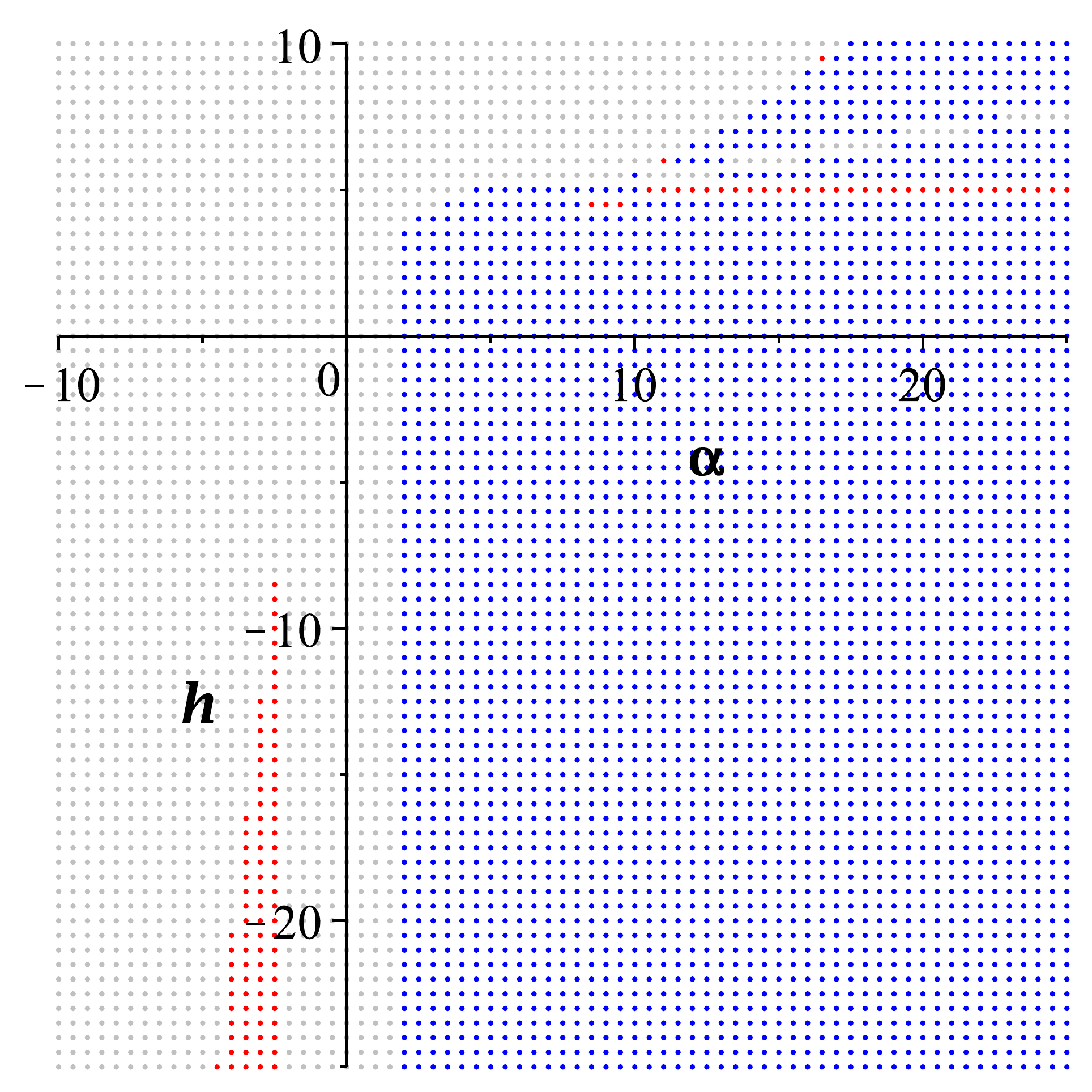}
\includegraphics[width=0.3\textwidth]{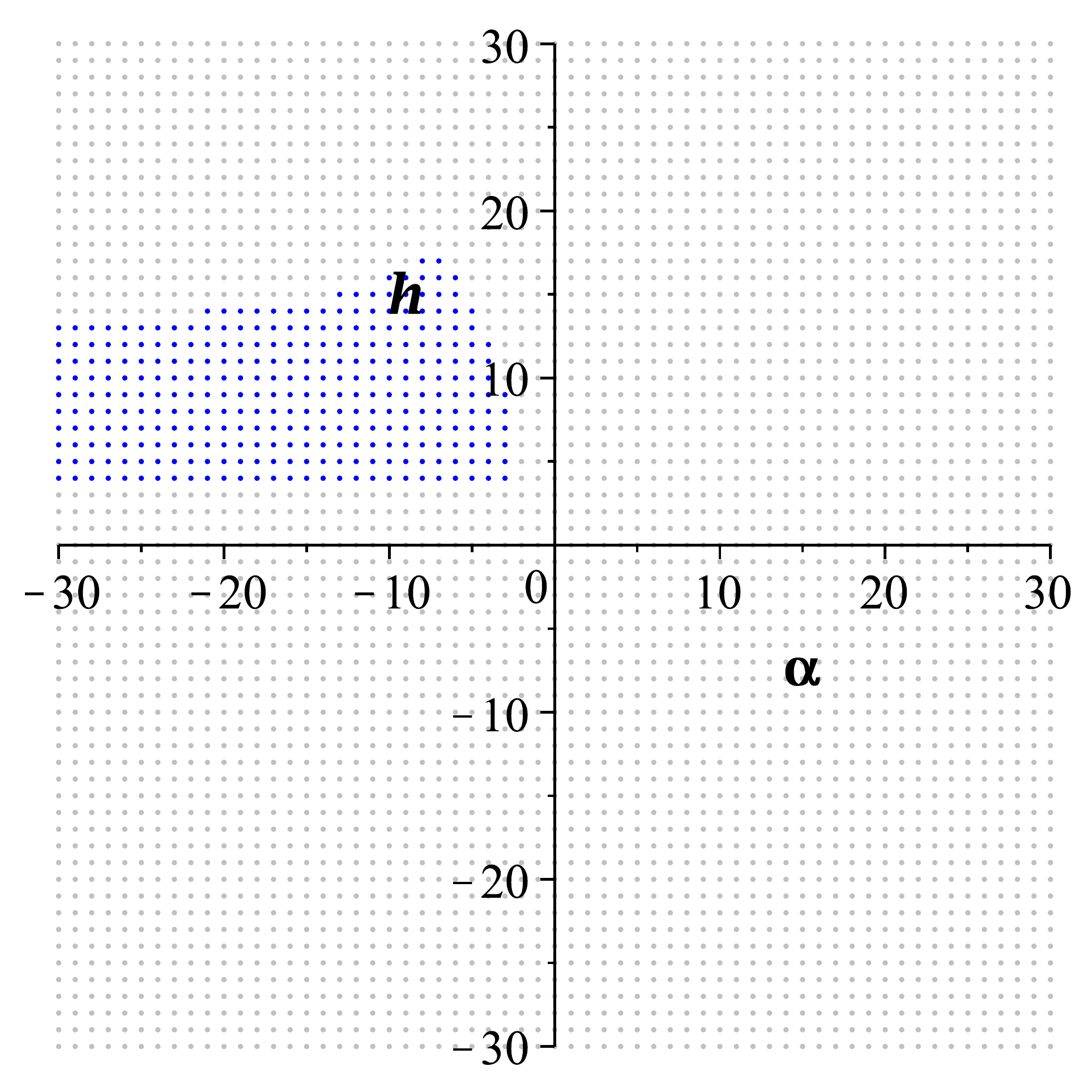}
\caption{{\bf Possible critical points in $(\alpha,\thh)$ parameter space with charge}:  \textit{Left}: for $k = 1$, $\epsilon = +1$ and $q = 2$.   \textit{Right}: for $k = -1$, $\epsilon = +1$ and $q = 2$.   Possible critical points satisfying entropy and pressure conditions, where red, blue, and grey points indicate two, one, and zero possible critical points, respectively.  In each case, only critical points which satisfy the various physicality constraints and minimize the Gibbs free energy have been included. }
\label{fig:pos_mu_number_critical_points_k=1_charge}
\end{figure}

Moving to the case where charge is present, completely scanning the parameter space becomes much more difficult since it is now three dimensional.  Therefore we provide here only a representative sampling of the thermodynamic phase space. First, in the spherical case, adding electric charge has simple effects on the thermodynamics.  The $\alpha-\thh$ space looks qualitatively the same as the uncharged case, as shown in the left plot of figure~\ref{fig:pos_mu_number_critical_points_k=1_charge}, and we see no novel examples of critical behaviour beyond that already found in the charge-free case.  As the magnitude of the electric charge is increased, the qualitative structure of the critical point diagram and corresponding critical behaviour remains the same. 

When charge is included for hyperbolic black holes, a region containing a single physical critical point emerges (see figure~\ref{fig:pos_mu_number_critical_points_k=1_charge}).  These critical points correspond to the end point of a first order van der Waals type phase transition for these black holes.

\subsubsection{Negative $\mu$ thermodynamics}

\begin{figure}[htp]
\centering
\includegraphics[width=0.3\textwidth]{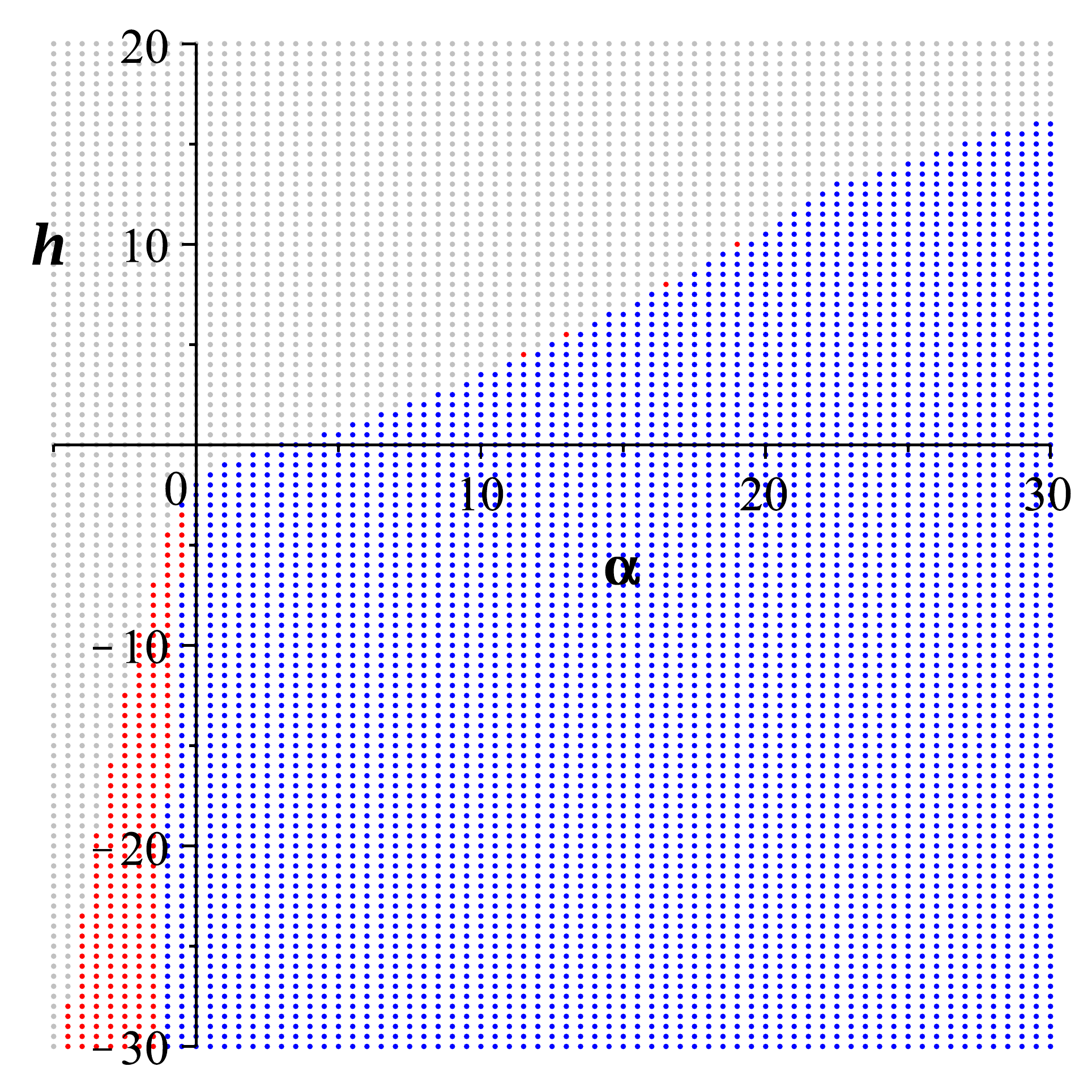}
\caption{{\bf Possible critical points in $(\alpha,\thh)$ parameter space for $k = 1$, $\epsilon=-1$ and $q = 0$}:   Possible critical points satisfying entropy and pressure conditions and minimizing the Gibbs free energy.  Note that there is a very narrow band where two critical points occur  between the boundary of the blue and grey regions in the upper-right quadrant, appearing as a few isolated red dots in the plot. Here red, blue, and grey points indicate two, one, and zero possible critical points respectively.}
\label{fig:neg_mu_number_critical_points_k=1}
\end{figure}

We next consider the case of negative quasi-topological coupling ($\epsilon=-1$).  Considering first the case of uncharged, spherical black holes we find the space of possible critical points is of the form shown in figure~\ref{fig:neg_mu_number_critical_points_k=1}.  Note that, while not hard to capture in this figure, there is a narrow band  (illustrated by a few red dots on the boundary between the blue and grey regions in the upper-right quadrant)
between the blue and grey region in the upper-right quadrant where two physical critical points are possible. For much of the parameter space we see the same type of behaviour as observed in the positive coupling case.  For example, we see van der Waals behaviour in the lower left quadrant (e.g for $\alpha = -2$ with $\thh = -20$), and re-entrant phase transitions occur in the blue/red region in the upper right quadrant (e.g. at $\alpha=25$ with $\thh = 10$).  

\begin{figure}[htp]
\centering
\includegraphics[width=0.3\textwidth]{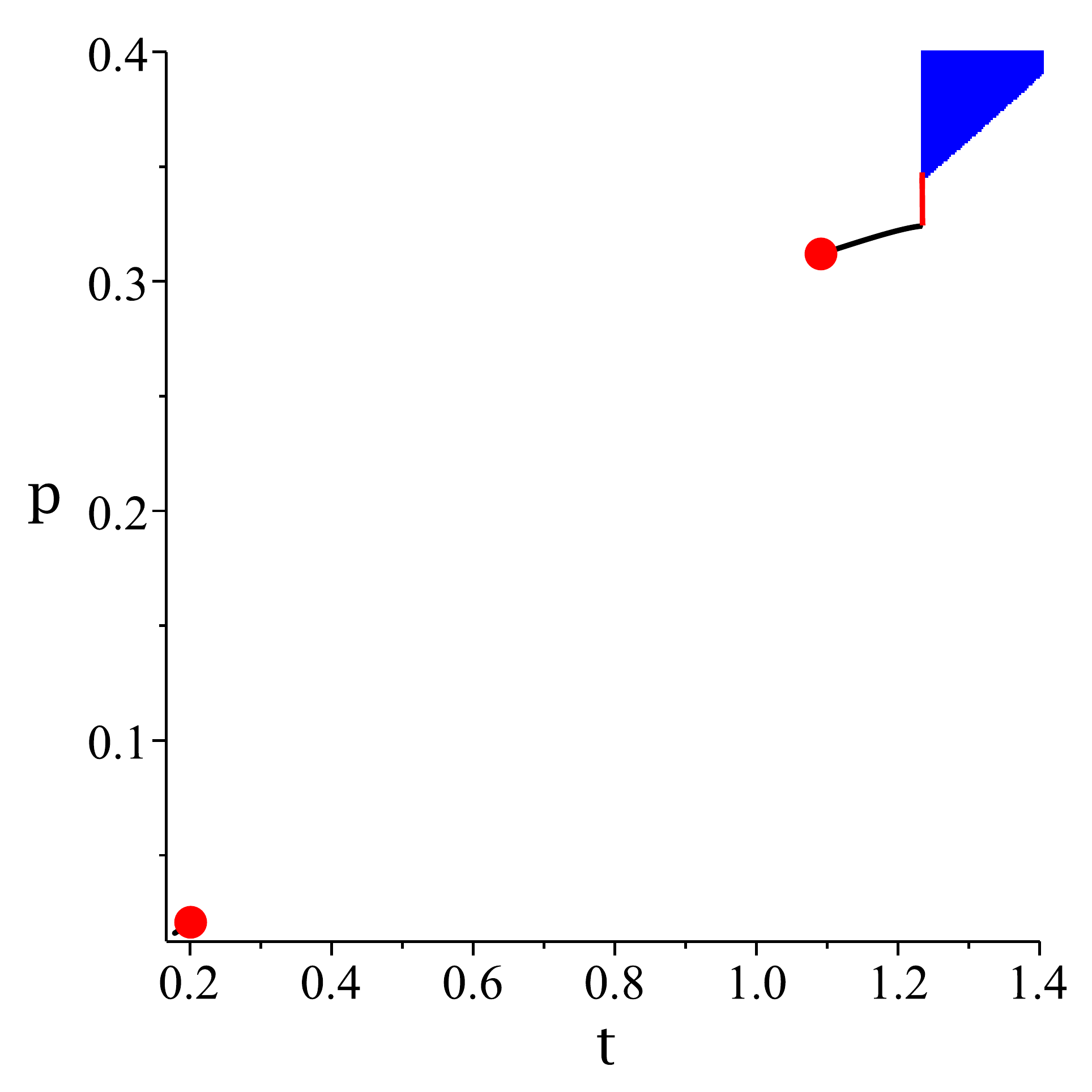}
\includegraphics[width=0.3\textwidth]{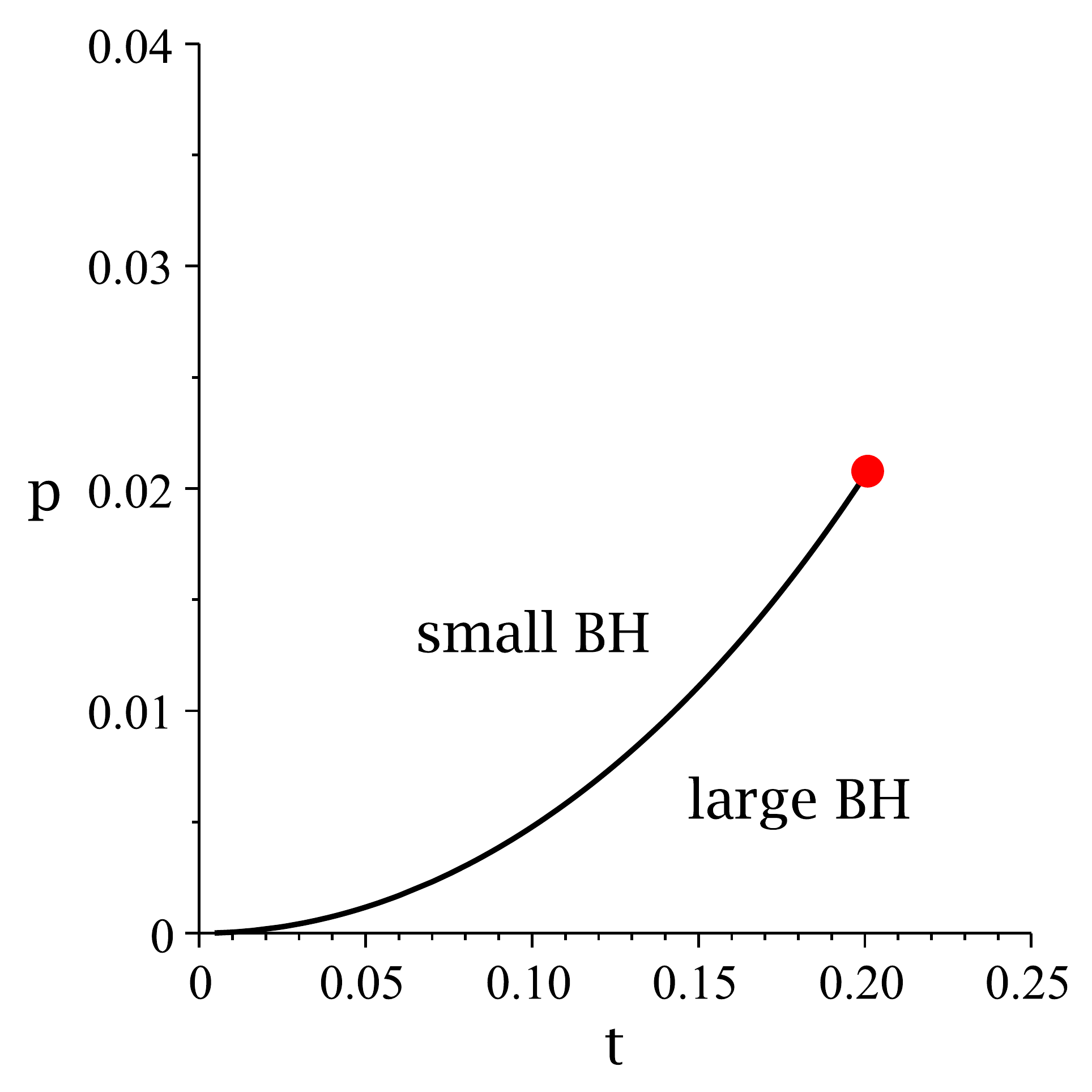}
\includegraphics[width=0.3\textwidth]{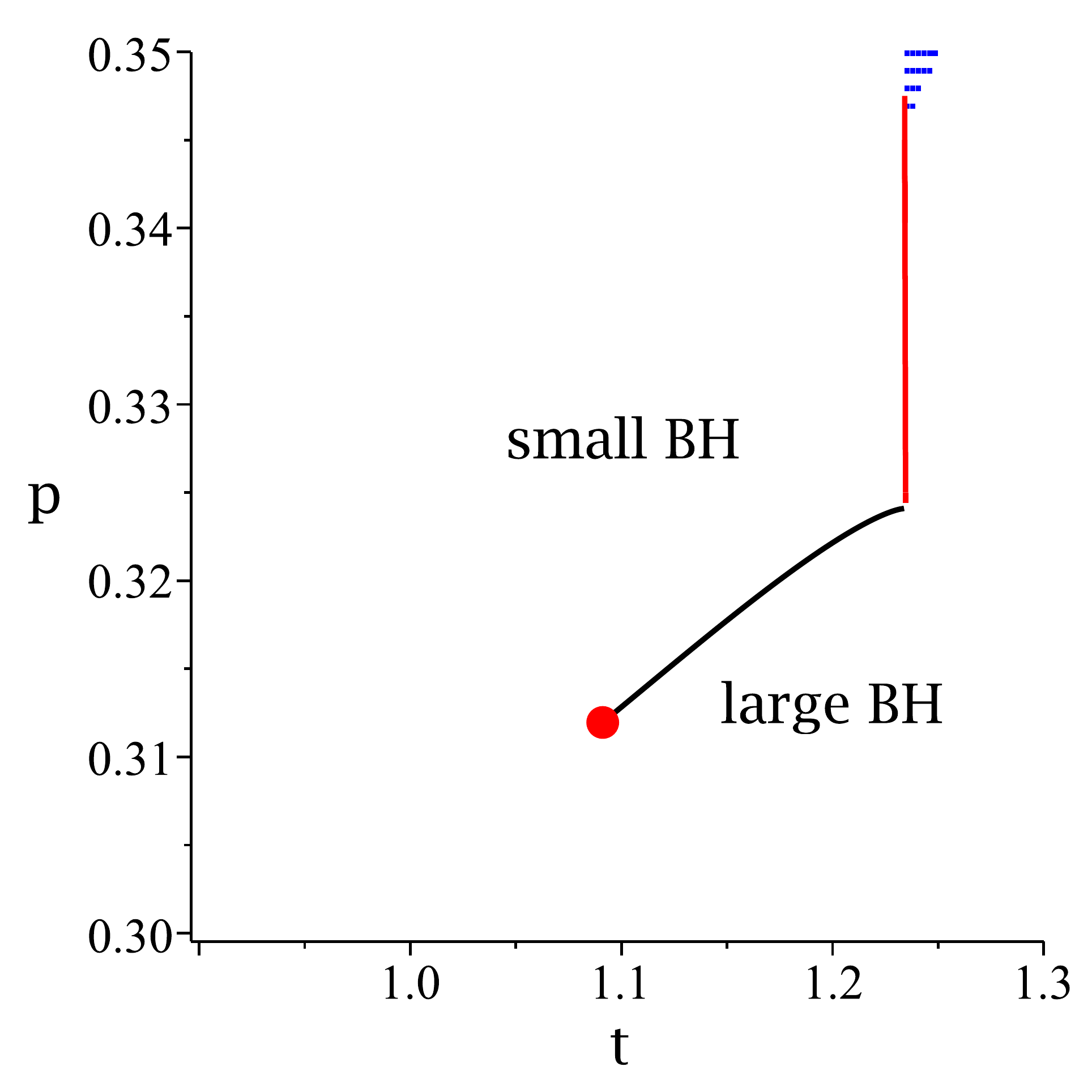}
\caption{{\bf Phase diagram for $k = +1$, $\epsilon=-1$, $\alpha = -0.6$, $\thh = -5$ and $\tq = 0$}: \textit{Left}: The full phase diagram.  \textit{Center}: A zoomed-in view of the lower-left portion of the phase diagram, showing VdW behaviour.  \textit{Right}: A zoomed-in version of the upper-right portion of the phase diagram, showing zeroth and first order phase transitions. The large red circles correspond to critical points, the solid black line is the coexistence line of a first order phase transition, and the solid red line marks a zeroth order phase transition. In the blue shaded region, no black hole solutions exist.}
\label{fig:novel_phase_gap}
\end{figure}

Characteristic behaviour of the two-critical-point case is shown in figure~\ref{fig:novel_phase_gap}. This is an interesting phase diagram of a qualitative kind not seen before.  Here we see typical van der Waals behaviour for temperatures and pressures in the lower left corner of the phase diagram (enlarged in the central part of  figure~\ref{fig:novel_phase_gap}).  In the upper right of the phase diagram (enlarged in the right-hand part of  figure~\ref{fig:novel_phase_gap}) we see a region where  both zeroth and first order phase transitions can occur: a line of first order phase transitions (black line) stems from a critical point (red dot) and terminates where it meets with a line of zeroth order phase transitions (red line).  The line of zeroth order phase transitions ends at the beginning of a ``no black hole region" (shaded blue area) where no black hole solutions exist; in this region the Gibbs free energy appears as two disjoint cusps.  In the region between the two critical points there is no notion of distinct phases.

\begin{figure}[htp]
\centering
\includegraphics[width=0.4\textwidth]{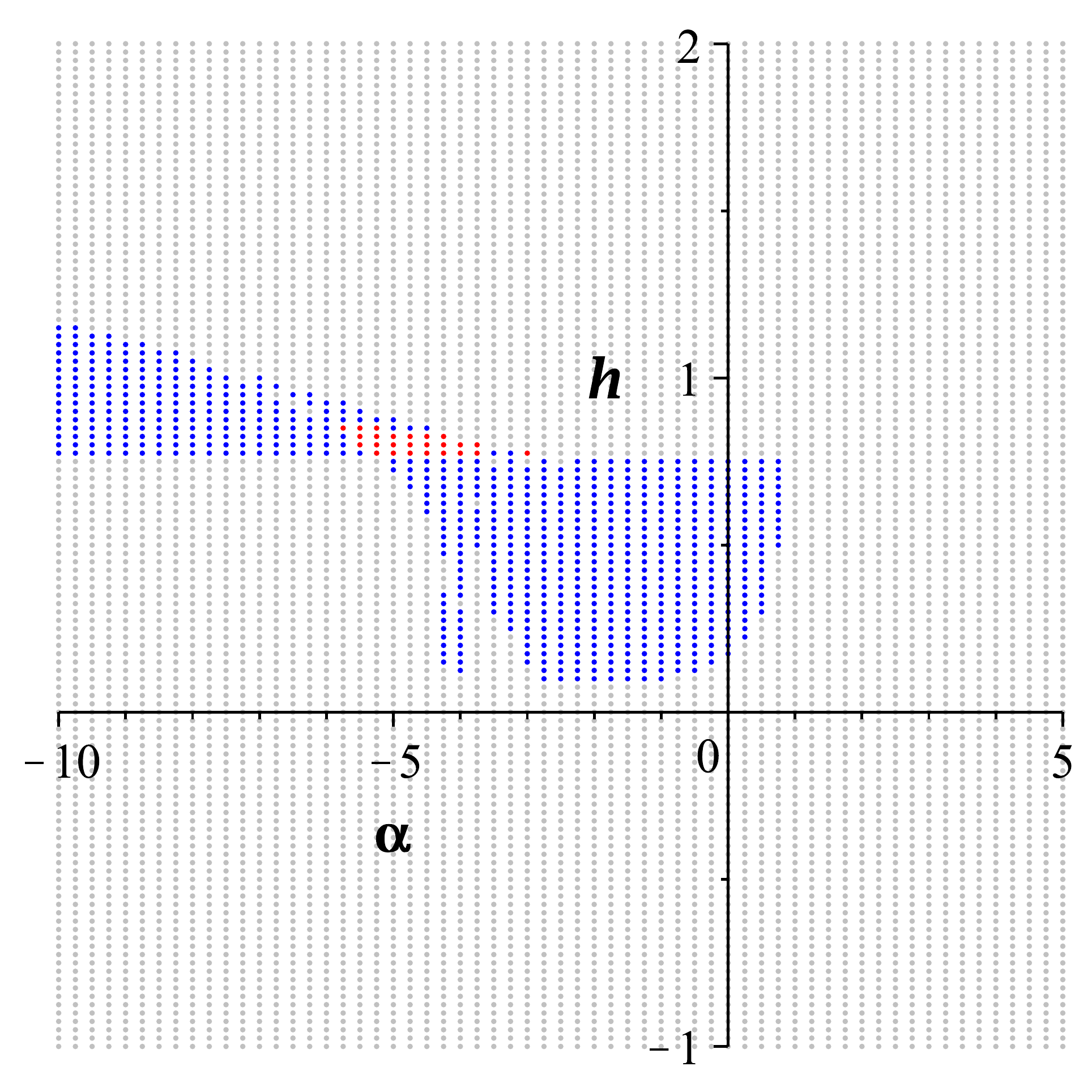}
\includegraphics[width=0.4\textwidth]{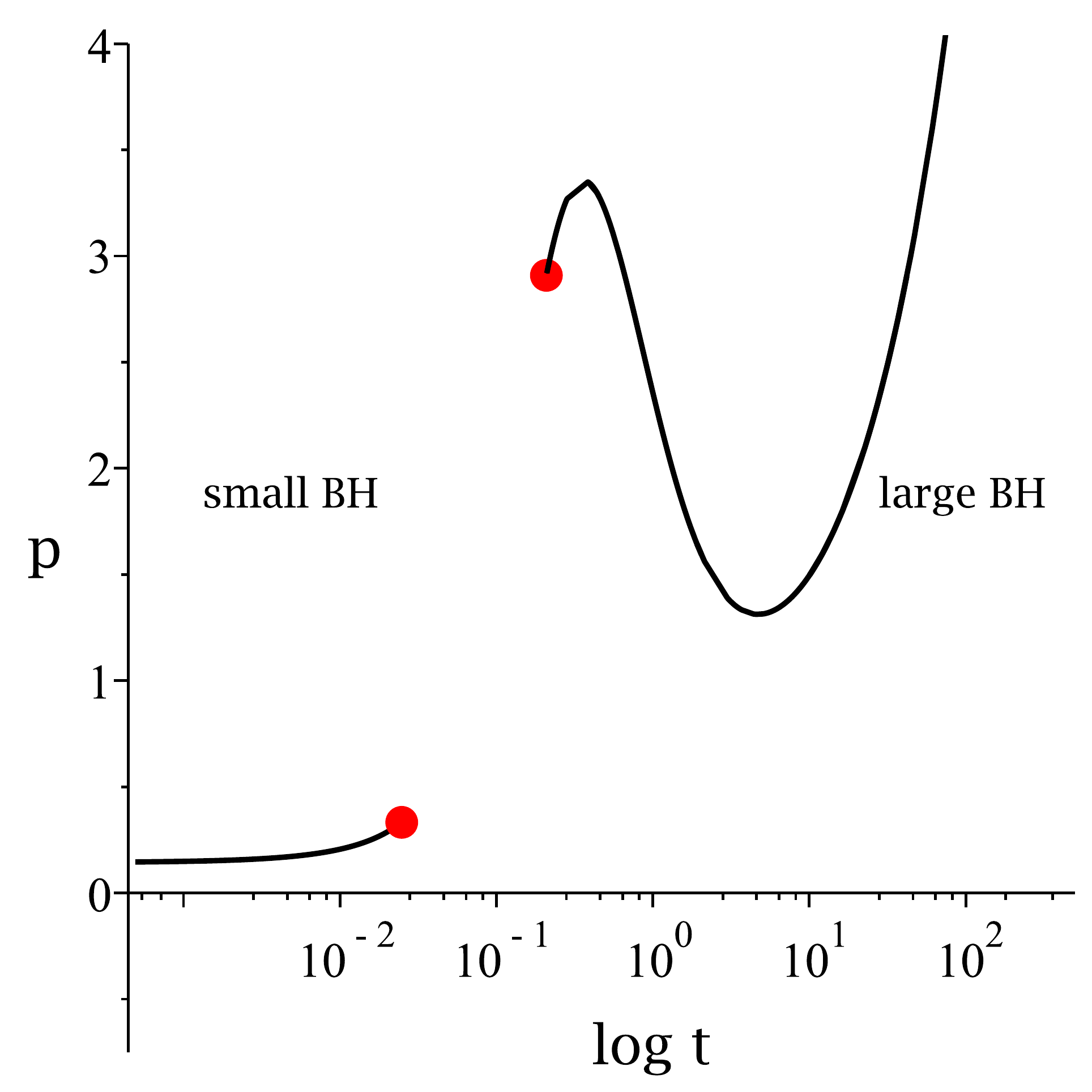}

\caption{{\bf Hyperbolic black holes without charge}: \textit{Left}: Critical point profile for $\epsilon=-1$, $k=-1$ with $q=0$.  In producing this plot, the only points included are those which satisfy all physicality conditions and also occur at a minimum of the Gibbs free energy. Here red, blue, and grey points indicate two, one, and zero critical points, respectively. \textit{Right}: Representative phase behaviour for the region with two critical points, shown here for $\alpha = -5$ with $\thh = 0.825$.  Here the two solid, red dots indicate critical points, while the solid black curves mark first order coexistence lines.  Here we see an example of a small/large/small/large black hole reentrant phase transition for $p$ in the approximate range $2.8$ to $3.3$ as temperature is increased monotonically.}
\label{fig:crits_no_charge_neg_mu_hyper}
\end{figure}

In the case of hyperbolic black holes without charge, we find that the critical behaviour is more interesting than the positive coupling case, as can be seen from the critical point profile shown in figure~\ref{fig:crits_no_charge_neg_mu_hyper}. The new feature here is the presence of a narrow region where up to two critical points can occur.  In the case of single critical point, we see either van der Waals or reverse van der Waals behaviour, but an interesting phase diagram occurs for the case of two critical points.  Representative behaviour is shown in figure~\ref{fig:crits_no_charge_neg_mu_hyper} where we note that this novel phase diagram results in a small/large/small/large black hole reentrant phase transition at sufficiently high pressure   (roughly, for $p$ between 2.8 and 3.3 in the right plot of figure~\ref{fig:crits_no_charge_neg_mu_hyper}). When $\thh$ is chosen carefully, the two critical points here can coincide to give an isolated critical point combined with a multiple re-entrant phase transition.  We defer a more detailed discussion of isolated critical points to section~\ref{sec:icp_section}, but refer the reader in particular to figure~\ref{fig:icp_hyper}.

\begin{figure}[htp]
\centering
\includegraphics[width=0.4\textwidth]{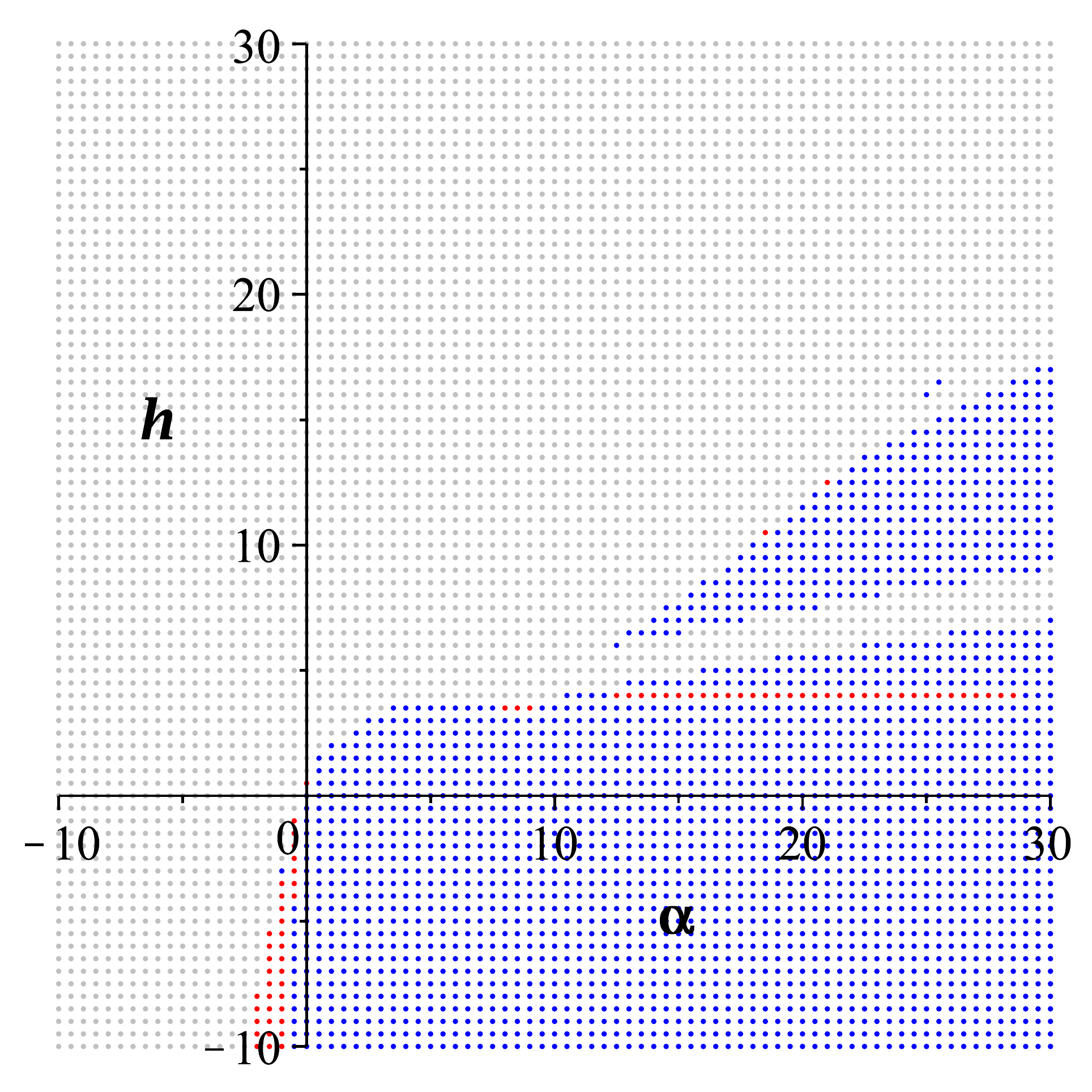}
\includegraphics[width=0.4\textwidth]{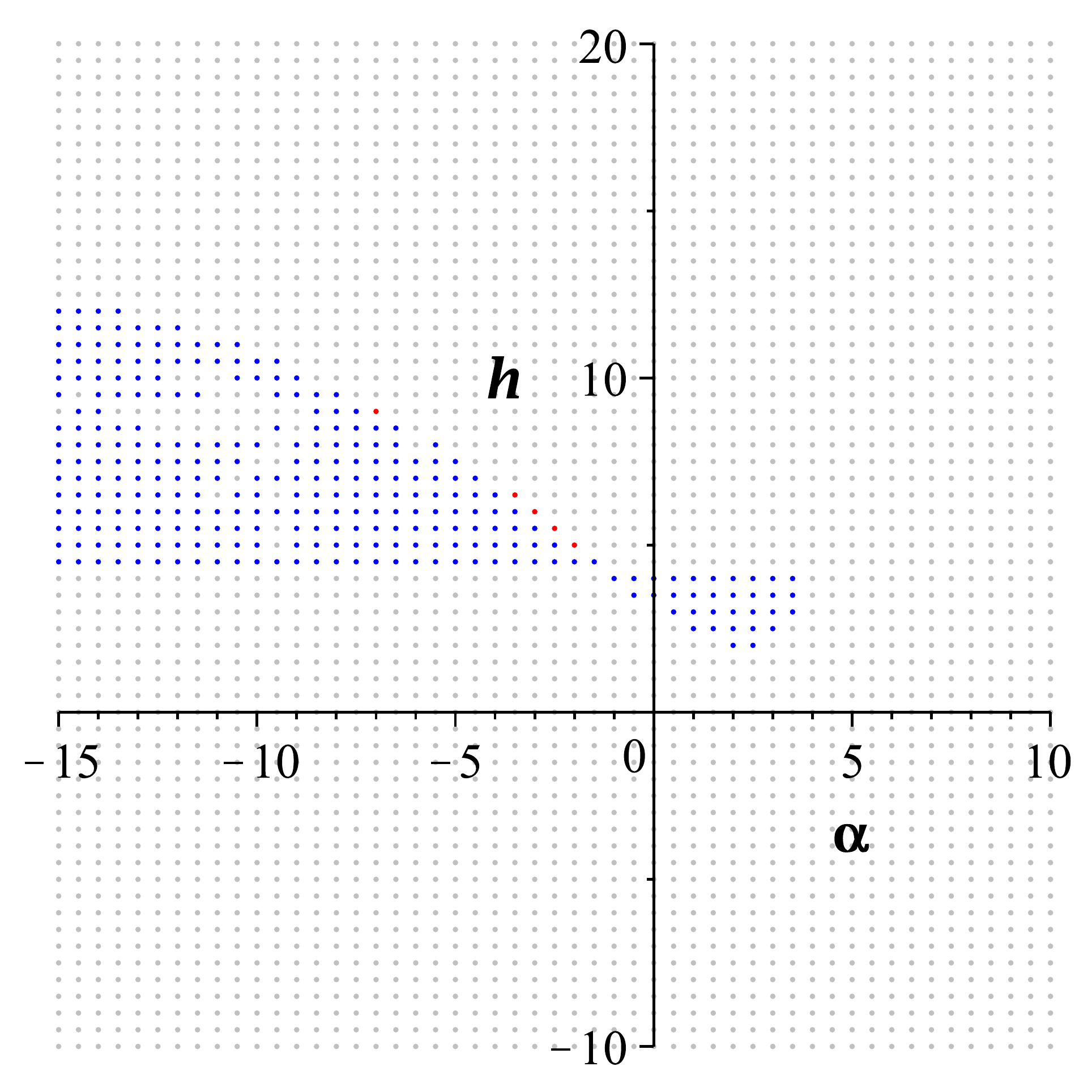}
\caption{{\bf Critical point profiles for charged black holes ($\tq = 2$)}:  The case for negative quasi-topological coupling.  In each plot, red, blue, and grey points 
indicate two, one, and zero critical points, respectively. \textit{Left}: Spherical black holes.  \textit{Right}: Hyperbolic black holes.}
\label{fig:negative_coupling_with_charge}
\end{figure}

Our last consideration involves the inclusion of electric charge in the negative coupling case.  Once again, this results in a large parameter space that cannot be (reasonably) fully explored.  However, after an intensive investigation of the parameter space, we have found that the general behaviour provides nothing beyond the examples we have already discussed up to this point. For both the spherical and hyperbolic cases, a representative plot of the critical point space is shown in figure~\ref{fig:negative_coupling_with_charge}.  The left plot of this figure is qualitatively similar to figure~\ref{fig:neg_mu_number_critical_points_k=1}, and the critical behaviour is analogous.  A similar story holds for the right plot of figure~\ref{fig:neg_mu_number_critical_points_k=1} which shows a typical critical point profile in the charged hyperbolic case.  This profile resembles that shown in figure~\ref{fig:crits_no_charge_neg_mu_hyper} for the uncharged hyperbolic black holes, and we find analogous critical behaviour.  The general trend is that, as the value of the charge is increased, the region of parameter space with two physical critical points shrinks, eventually vanishing.

\subsection{Superfluid black hole behaviour}

In~\cite{Hennigar:2016xwd} a black hole $\lambda$-line was observed for a class of Lovelock black holes  possessing the same conformal hair considered here.  This line of second order phase transitions bears resemblance to the fluid/superfluid transition in liquid $^4$He, and was found to occur in all dimensions $D \ge 7$.  In that work, a set of precise conditions were given that are necessary for a black hole equation of state to satisfy to give rise to the $\lambda$-line.  Here, we explore this phenomenon for the five-dimensional quasi-topological black holes.

As we have seen, the equation of state of these quasi-topological black holes takes the form,
\be 
p = \frac{t}{v} - \frac{3k}{2 \pi v^2} + \frac{2 \alpha k t}{v^3} + \epsilon \frac{3 k^2 t}{v^5} + \epsilon \frac{3 k^3}{2 \pi v^6} + \frac{\tq^2}{v^6} - \frac{\thh}{v^5} \, ,
\ee
where $\epsilon = {\rm sign}(\mu)$ represents the sign of the quasi-topological coupling.  This equation of state has the form of eq.~(18) in \cite{Hennigar:2016xwd} with 
\be
a_1 = \frac{1}{v} + \frac{2\alpha k}{v^3} + \epsilon \frac{3 k^2}{v^5} \, , \quad a_2 = -\frac{3 k}{2 \pi v^2} + \epsilon \frac{3 k^3}{2 \pi v^6} - \frac{\thh}{v^5} + \frac{\tq^2}{v^6} \, .
\ee
The condition for a black hole $\lambda$-line is given by the simultaneous solution of
\be 
\frac{\partial a_i}{\partial v} = 0 \, , \quad \frac{\partial^2 a_i}{\partial v^2 } = 0 \,  \text{ for } i = 1, 2  
\ee
which amounts to solving the conditions for a critical point without placing any restrictions on what $t_c$ should be.  Here we find that a solution to this system exists, but only for the following parameters:
\begin{align}
\epsilon &= +1 \, , \quad k =-1 \, , \quad  \alpha  = \sqrt{\frac{5}{3}} \, , \quad v_c = 15^{1/4} \, , \quad \thh =  \frac{12 (15)^{3/4}}{5 \pi} \, , \quad \tq^2 = \frac{24}{\pi} \, .
\end{align}
The value of $\alpha$ above agrees with the result obtained by setting $d=5$ in the expressions from~\cite{Hennigar:2016xwd}, but the values of $\tq$ and $h$ are different.  This can be understood by realizing that in $D > 5$ there is an additional term which will appear in the equation of state (cf. eq.~(12) from \cite{Hennigar:2016xwd}).  

\begin{figure}[htp]
\includegraphics[width=0.3\textwidth]{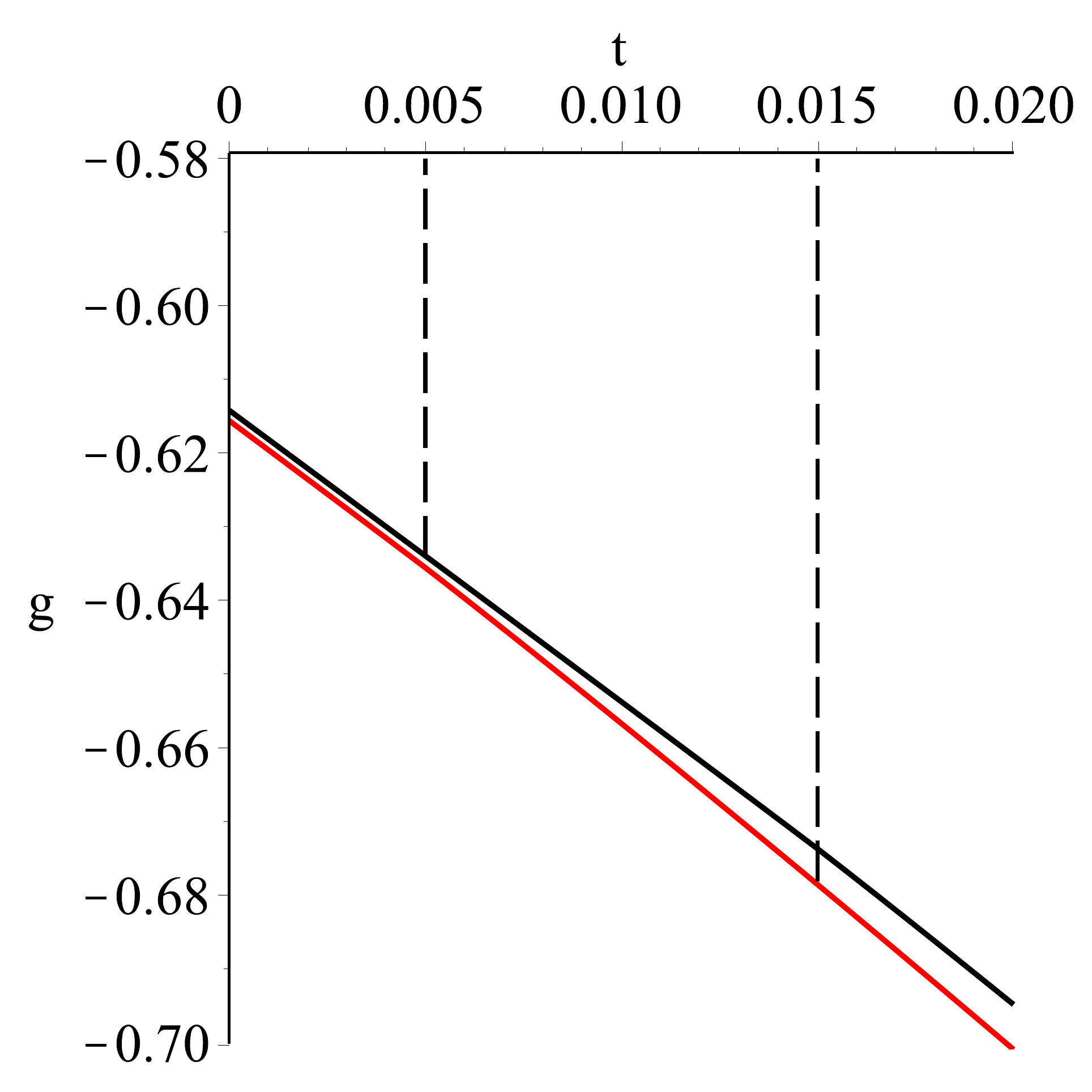}
\includegraphics[width=0.3\textwidth]{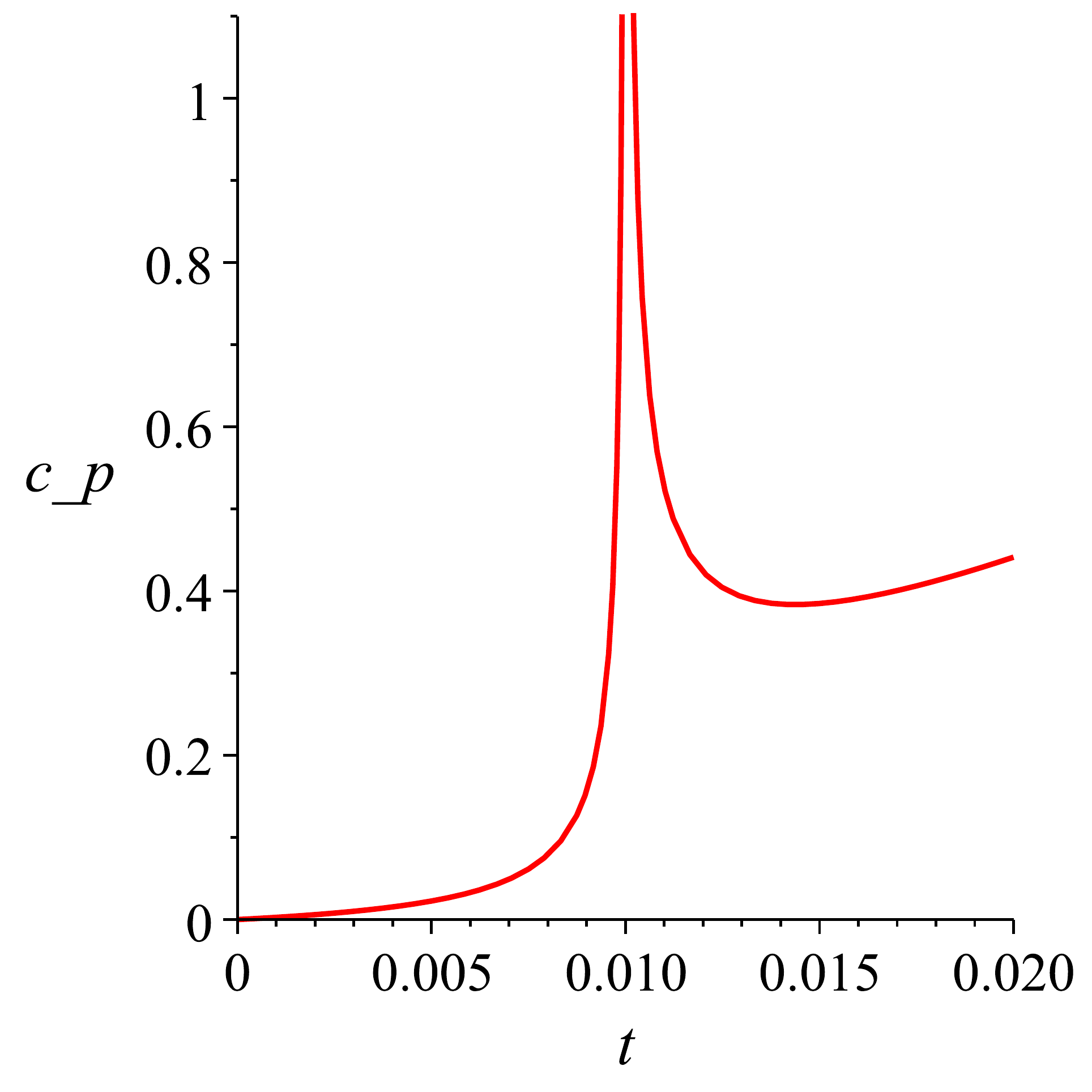}
\includegraphics[width=0.3\textwidth]{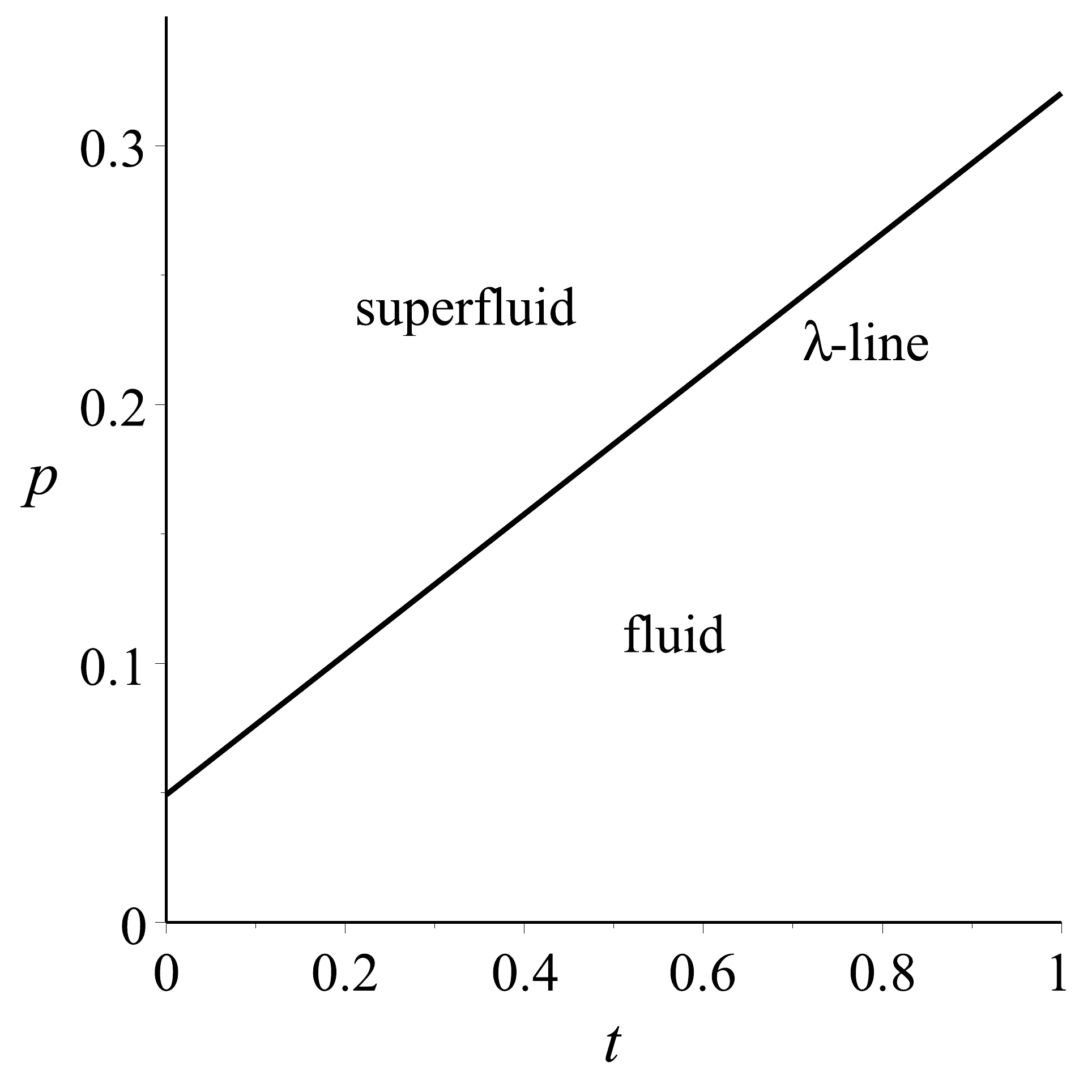}
\caption{{\bf Black hole $\lambda$-line}: {\it Left}: A plot of the gibbs free energy for two different values of pressure corresponding to $t_c = 0.005$ and $0.015$ as indicated by the vertical, dashed lines. At these points, the Gibbs free energy has a cusp, indicating a diverging specific heat. {\it Center}: A plot of the specific heat for the case where the second order phase transition occurs at $t_c = 0.01$, the shape is characteristic of the $\lambda$-line. {\it Right}: A plot of the phase diagram.  The solid black line is the $\lambda$-line, which corresponds to a line of second order phase transitions.}
\label{fig:superfluid_behaviour}
\end{figure}

For the above parameter values we have a line of critical points with the critical values,
\be 
v_c =  15^{1/4} \, , \quad p_c = \frac{8 (15)^{3/4}}{225} t_c + \frac{\sqrt{15}}{25 \pi} \, , \quad t_c \in \mathbb{R}^+ \, .
\ee
We emphasize that there is no first order phase transition associated with this line of critical points: there is simply a line of second order (continuous) phase transitions, as shown in figure~\ref{fig:superfluid_behaviour}.  Similar results arise in condensed matter physics where these lines are termed $\lambda$-lines due to the shape of the specific heat curve.  The presence of $\lambda$-lines often indicates the onset of some quantum phenomenon, such as fluid/superfluid transitions~\cite{RevModPhys.73.1}, superconductivity~\cite{superconductor}, and paramagentism/ferromagnetism transtions~\cite{Pathria2011401}.  Here the analogy with superfluidity is most appropriate since the thermodynamic analogy of black holes with fluids has considerable support in its favour~\cite{Kubiznak:2016qmn}.

The entropy of the black holes which possess this superfluid-like transition is positive.  Furthermore, since these black holes are five dimensional, the boundary dual theory would be four-dimensional.  As a result it should be possible to explore the holographic consequences of the black hole $\lambda$-line for four dimensional CFTs as an extension of existing literature~\cite{Myers:2010jv}.  We hope to return to this in the future. 

\subsection{Isolated critical points}
\label{sec:icp_section}

Here we consider the possibility of isolated critical points for these hairy quasi-topological black holes. These are critical points where the critical exponents do not match the mean field theory values.  Thus far, there have been very few such examples in the literature, with the first such examples found in the context of Lovelock and quasi-topological gravity~\cite{Hennigar:2015esa,Frassino:2014pha,Dolan:2014vba}.  In these cases it was found that the black holes possessing these isolated critical points required finely tuned coupling constants, hyperbolic horizons and occur only when the black hole is massless.  
In the case of Lovelock black holes with conformal scalar hair~\cite{EricksonRobie}, a family of isolated critical points were found in seven and higher dimensions.  This discovery provided the first example of isolated critical points occurring for black holes with freedom in the coupling parameters and for black holes of any mass.  Due to the similarities of the theories, we expect to see similar results here for the quasi-topological black holes with scalar hair.

Recall that the equation of state for these black holes is given by,
\be 
p = \frac{t}{v} - \frac{3k}{2 \pi v^2} + \frac{2 \alpha k t}{v^3} + \epsilon \frac{3 k^2 t}{v^5} + \epsilon \frac{3 k^3}{2 \pi v^6} + \frac{\tq^2}{v^6} - \frac{\thh}{v^5} \, ,
\ee
and that the condition for a critical point is that
\be\label{eqn:crit_point} 
\frac{\partial p}{\partial v} = 0 = \frac{\partial^2 p}{\partial v^2} \, .
\ee
Isolating these expressions for the temperature, we find
\be\label{eqn:crit_temp} 
t_c = \frac{3k v^4 + 5\pi \thh v - 6 \pi  \tq^2 - 9  \epsilon k^3}{\pi v \left( v^4 + 6 \alpha k v^2 + 15 \epsilon k^2 \right)} \, .
\ee
Clearly this expression appears to be ill-defined when the denominator vanishes, that is when,
\be\label{eqn:icp_volumes} 
v_{c, \delta, \epsilon} = \sqrt{k \left[-3\alpha + \delta \sqrt{9 \alpha^2 - 15 \epsilon} \right]} 
\ee
where $\delta = \pm 1$.  However, we can choose a value of $\thh$ such that the critical temperature is non-singular for these volumes;  explicitly, this is
\be 
\thh_{\delta, \epsilon} = -\frac{6}{5 \pi} \frac{9(\alpha^2-\epsilon)k^3 -  \pi \tq^2  - 3\alpha \delta k^3 \sqrt{9 \alpha^2 - 15 \epsilon}}{\sqrt{-3 \alpha k + \delta k \sqrt{9 \alpha^2 - 15 \epsilon}}} \, ,
\ee
where the charge remains a free parameter.  For this set up the conditions for a critical point are satisfied, with $v_{c, \delta, \epsilon}$ being the critical volume.   Remarkably, one can show that the values $v_{c, \delta, \epsilon}$ correspond to the coalescence of two (or more) critical points.  More precisely, this means that the equation,
\begin{align} 
 &\left( 30 \pi v_c ^{4} +108\pi\alpha k v_c^{2}+90 \pi \epsilon
{k}^{2} \right) \tq^{2}-3 k v_c^{8}+18 {k}^{2} \alpha v_c ^{6}+180 {k
}^{3} \epsilon v_c^{4}-20 \thh_{\delta, \epsilon} \pi v_c^{5} 
\nn\\
&-60\thh_{\delta, \epsilon}\pi \alpha k v_c^{3}+162 \epsilon{k}^{4}\alpha v_c^{2}+135{k}^{5} = 0
\end{align}
[which follows from eq.~\eqref{eqn:crit_point} upon substituting $t=t_c$ from eq.~\eqref{eqn:crit_temp}]  has a double root for $v_c = v_{c, \delta, \epsilon}$.  The values of temperature and pressure at this critical point are given by
\begin{align}
t_c &= \frac{1}{2}\,{\frac {27\,{k}^{3}\alpha\,\delta\,\sqrt {9\,{\alpha}^{2}-15\,
\epsilon}+ \left(63\,\epsilon  -81\,{\alpha}^{2} \right) {k}^{3}-3 \pi\,{\tq
}^{2} }{\sqrt {9\,{\alpha}^{2}-15\,\epsilon}\sqrt {k\delta\,\sqrt {
9\,{\alpha}^{2}-15\,\epsilon}-3\,k\alpha}\pi \,{k}^{2}\delta\, \left( 
3\alpha-\,\delta\,\sqrt {9\,{\alpha}^{2}-15\,\epsilon} \right) }} \, ,
\nn\\
p_c &= -\frac{3}{5}\,{\frac { \left( -27\,{\alpha}^{2}+25\,\epsilon+9\,\alpha\,\delta
\,\sqrt {9\,{\alpha}^{2}-15\,\epsilon} \right) {\tq}^{2}}{ {k}^{3
}\delta\, \left( -3\,
\alpha+\delta\,\sqrt {9\,{\alpha}^{2}-15\,\epsilon} \right) ^{4}\sqrt {9\,{\alpha}^{2}-15\,\epsilon}}}
\nn\\
&-\frac{9}{5}\,{\frac {594\,
\epsilon\,{\alpha}^{2}-378\,{\alpha}^{4}-100\,{\epsilon}^{2}+126\,
\delta {\alpha}^{3} \,\sqrt {9\,{\alpha}^{2}-15\,\epsilon}-93 \epsilon \delta \alpha
\sqrt {9\,{\alpha}^{2}-15\,\epsilon}}{ \left( -3\,
\alpha+\delta\,\sqrt {9\,{\alpha}^{2}-15\,\epsilon} \right) ^{4}\pi \,
\delta\,\sqrt {9\,{\alpha}^{2}-15\,\epsilon}}} \, .
\end{align}
 
It is straightforward to show that near the critical point the equation of state is of the form,
\be\label{eqn:p-isocrit}
\frac{p}{p_c} = 1 + A \tau + B \omega^2 \tau + C \omega^3 + \cdots \, ,
\ee
where $\tau =(t-t_c)/t_c$ and $\omega=(v-v_c)/v_c$.  In this expansion, the explicit expressions for the constants $A, B$ and $C$  can be computed exactly, but the expressions are not enlightening. From this form of the equation of state we can conclude that the critical exponents are given by,
\be\label{eqn:cl_critexponent1}
\alpha=0,\, \, \beta=1,\, \, \gamma=2, \, \, \delta=3
\ee
which are non-standard critical exponents, but agree with those found in ~\cite{Dolan:2014vba}.

The first isolated critical points discovered \cite{Frassino:2014pha,Dolan:2014vba} were found to coincide with a thermodynamic singularity; we refer the reader to ref.~\cite{Frassino:2014pha} for a more detailed discussion of this phenomenon and only comment briefly here.  The thermodynamic singularity has somewhat undesirable features: for example, for all but a single value of pressure, there is a curvature singularity which occurs at the thermodynamic singularity.  In thermodynamic terms, it is a point in the parameter space where, in a $p-v$ diagram, all the isotherms intersect at a single point, i.e. this satisfies
\be 
\frac{\partial p}{\partial t} \Big |_{v_s} = 0\, .
\ee
The isolated critical points shown above coincide with the thermodynamic singularity only if
\be 
\frac{\partial p}{\partial t} \Big |_{v_{c,\delta,\epsilon}=v_s} = 0 \Rightarrow  v_{c,\delta,\epsilon}^{4}+2\alpha k v_{c,\delta,\epsilon}^{2}+3\epsilon{k}^{2}= 0
\ee
(cf. eq.~(3.19) in ref.~\cite{Frassino:2014pha}), and demanding that 
\be 
\alpha = \pm \sqrt{3 \epsilon}  
\ee
ensures that $v_{c,\delta,\epsilon}$ solves this equation.  However note  that this constraint is not necessary for the
isolated critical points obtained above, and therefore in general these do not coincide with a thermodynamic singularity.  Indeed,  the coefficient $A$ in the above near-critical expansion is generically non-zero.

Our discussion thus far has not focused on the circumstances under which these critical exponents are physical, or which---if any---constraints they violate.  We turn now to a discussion of these issues.  
\begin{figure}[htp]
\centering
\includegraphics[width=0.4\textwidth]{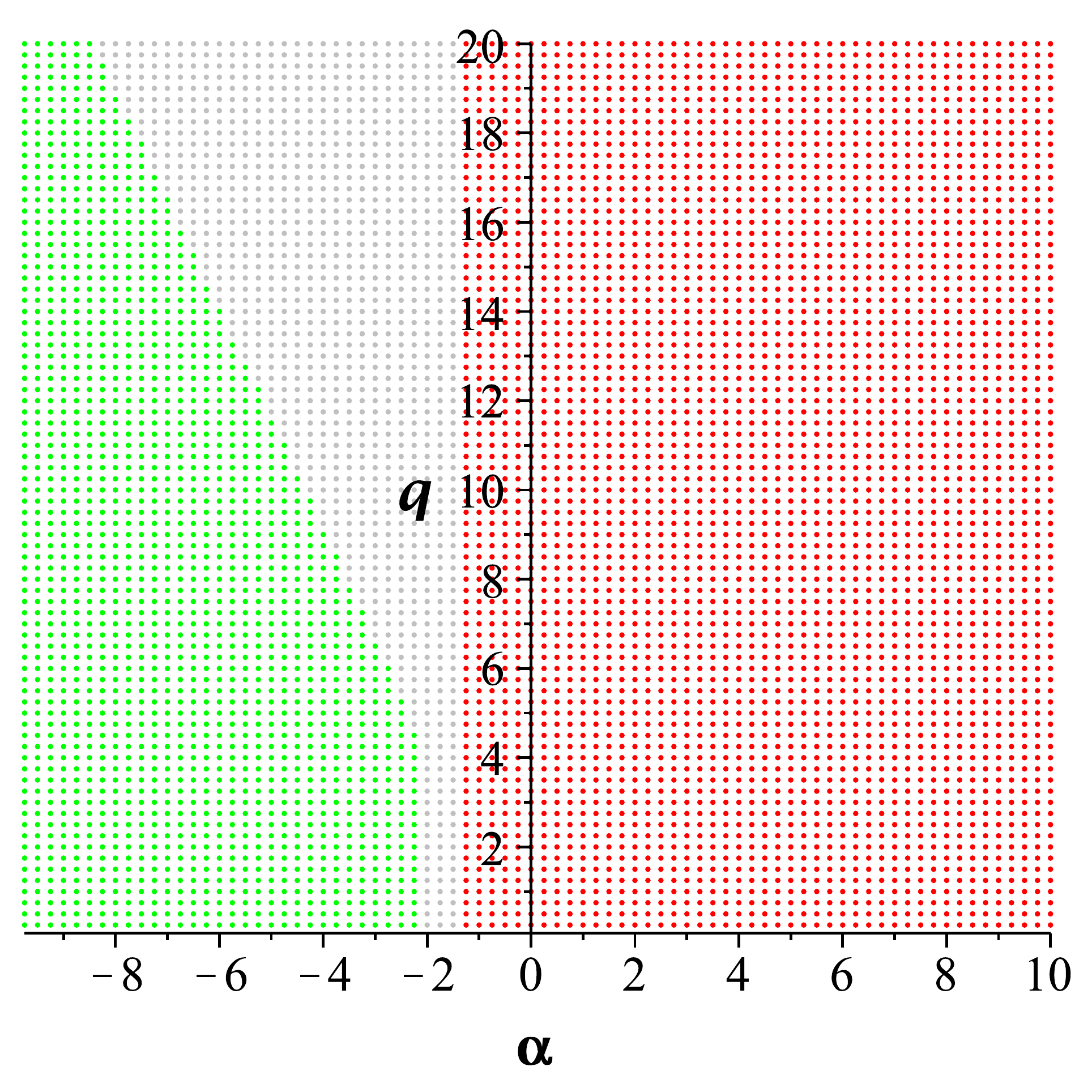}
\includegraphics[width=0.4\textwidth]{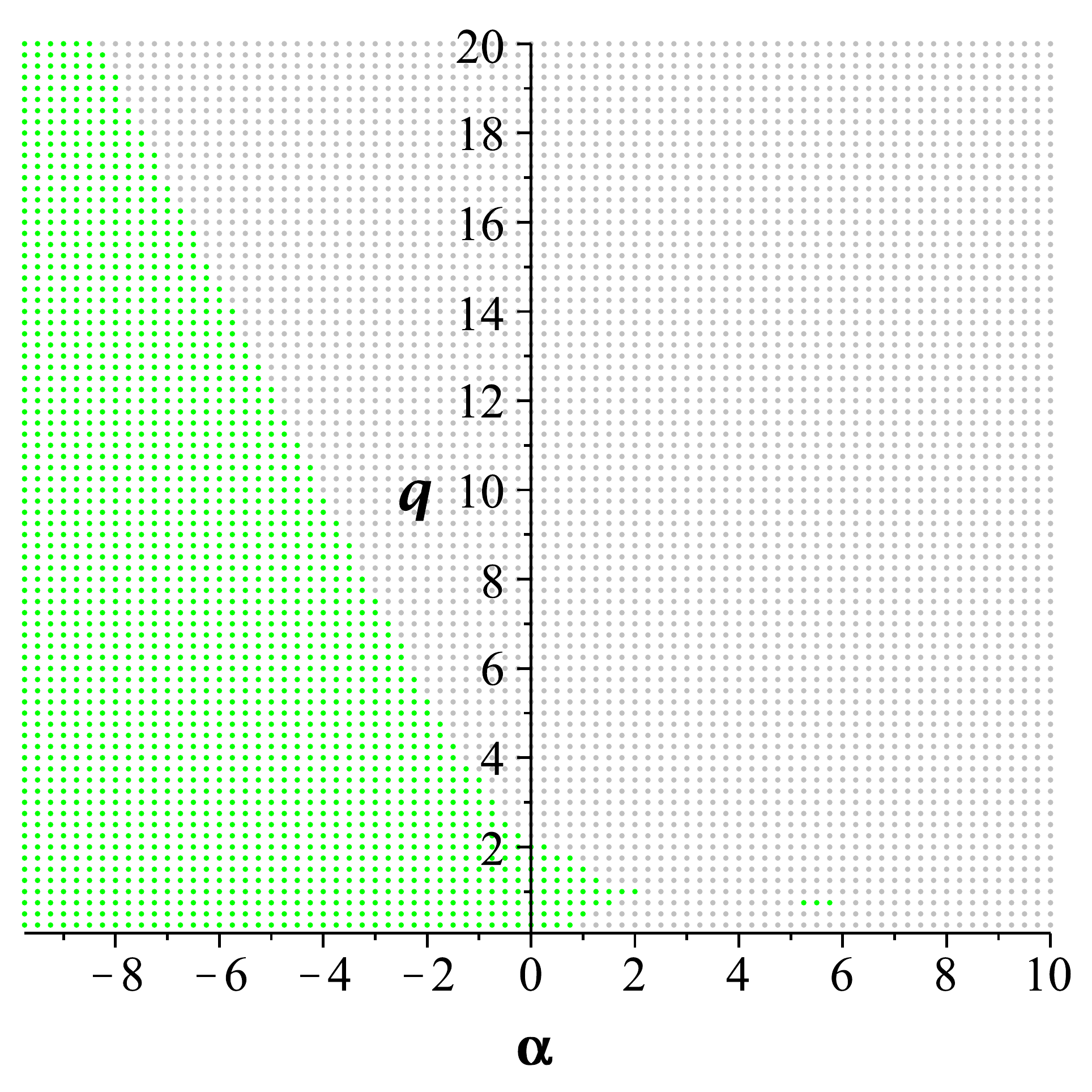}
\includegraphics[width=0.4\textwidth]{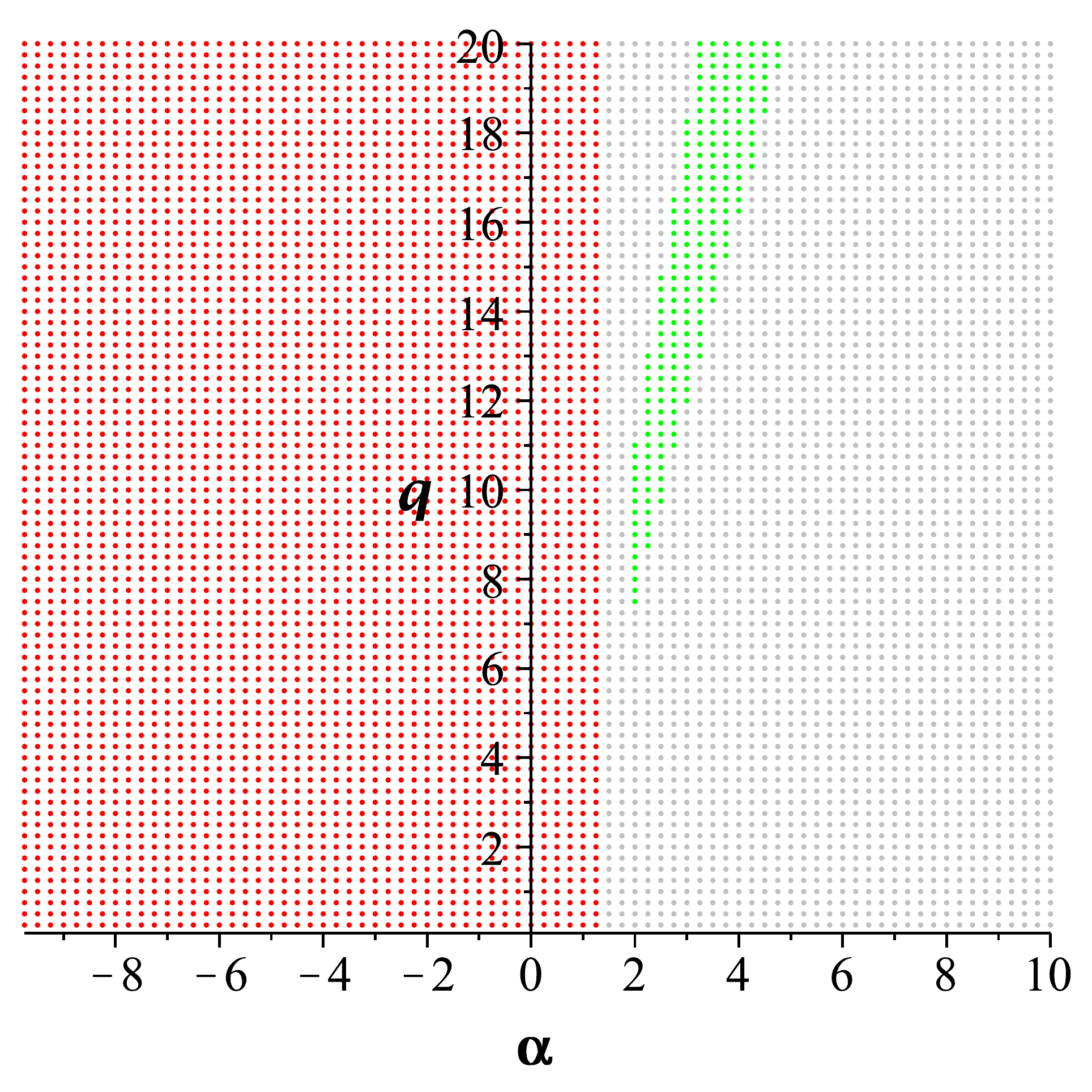}
\includegraphics[width=0.4\textwidth]{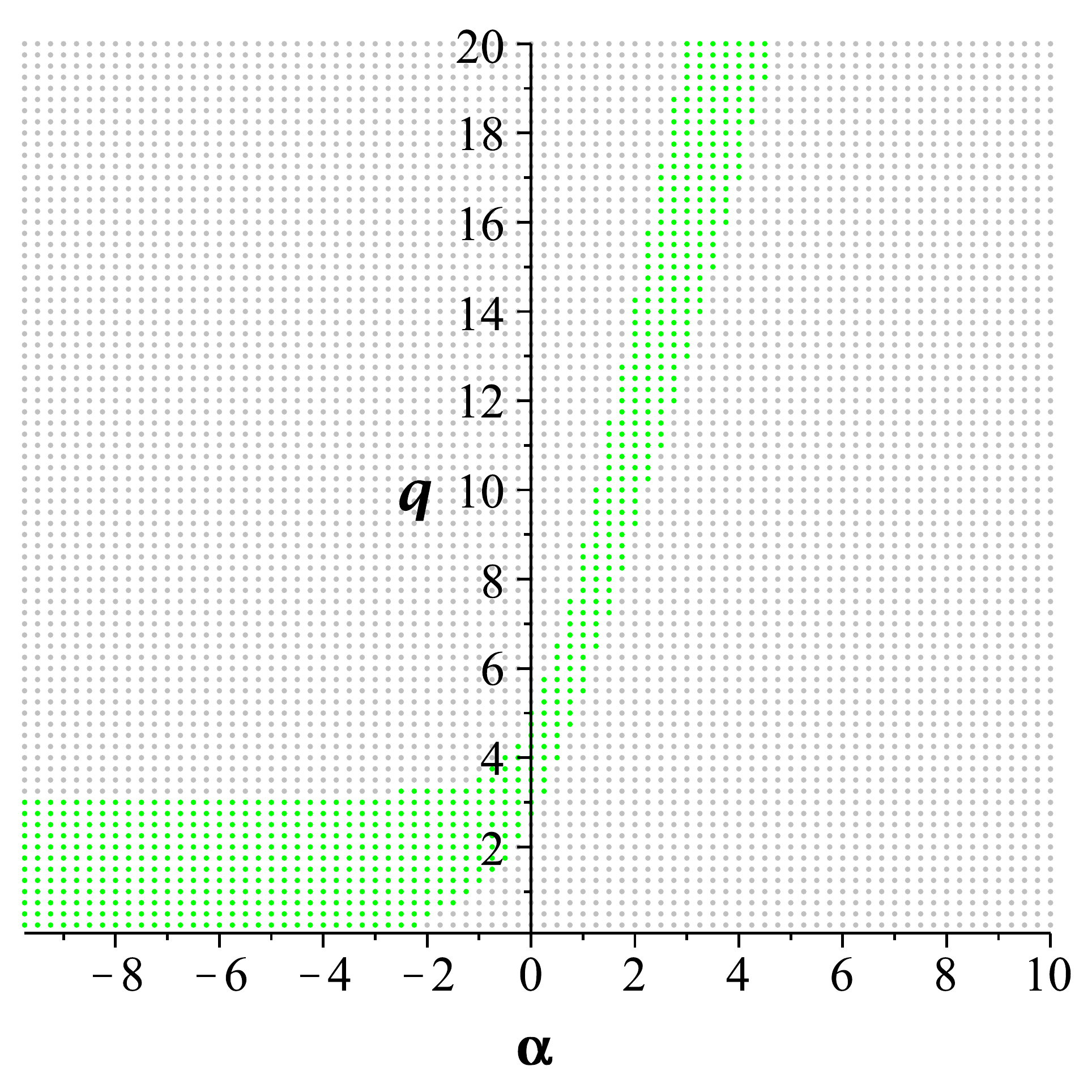}
\caption{{\bf Parameter space for isolated critical points}: {\it Top left}: Plot for $\delta = +1$, $k=+1$ and $\epsilon=+1$.  {\it Top right}: Plot for $\delta = +1$, $k=+1$ and $\epsilon=-1$. {\it Bottom left}: Plot for $\delta = -1$, $k=-1$ and $\epsilon=+1$.  {\it Bottom right}: Plot for $\delta = -1$, $k=-1$ and $\epsilon=-1$. In each plot, green dots indicate that for the selected parameters all thermodynamic parameters are positive and the (pressure and entropy) constraints are satisfied, grey dots indicate that for the chosen parameters at least one of the constraints is violated, red dots indicate that at least one of the thermodynamic parameters is imaginary.  All other combinations of $\delta, k$ and $\epsilon$ lead to situations where the physicality conditions cannot be satisfied.}
\label{fig:icp_parameter_space}
\end{figure}

We first consider the case for spherical black holes ($k=+1$).  In this case if $\delta = -1$, then there are no isolated critical points which satisfy all constraints for either sign of quasi-topological coupling.  However, interesting results are found for $\delta=+1$: as shown in figure~\ref{fig:icp_parameter_space}, for $
\delta = +1$, there exist large regions of parameter space where isolated critical points exist with all thermodynamic constraints satisfied.  These mark the first examples in the literature to date of isolated critical points occurring for black holes with spherical horizon topology. 

In the case of black holes with hyperbolic horizons ($k=-1$), there are also examples of isolated critical points.  This case is less surprising since all examples of isolated critical points found so far occur for black holes with hyperbolic horizons. In this case we find that, for $\delta = +1$, the valid solutions that occur for positive quasi-topological coupling violate the entropy constraint, while there are no valid solutions at all for negative quasi-topological coupling (the critical volume is imaginary for all $\alpha$).   In the case $\delta = -1$, there exist isolated critical points, regardless of the sign of the quasi-topological coupling, in certain regions of the parameter space, as shown in figure~\ref{fig:icp_parameter_space}. 
\begin{figure}
\centering
\includegraphics[width=0.3\textwidth]{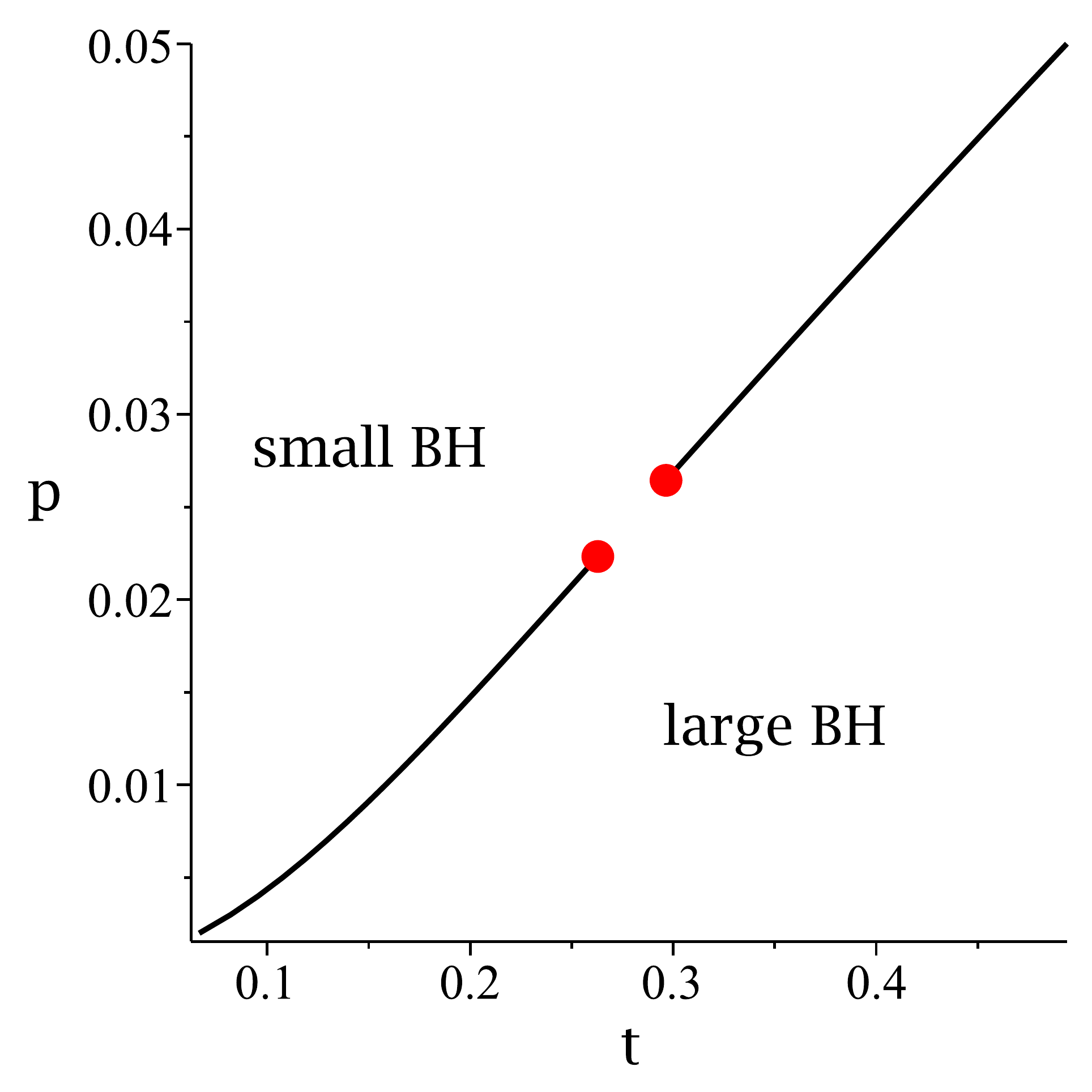}
\includegraphics[width=0.3\textwidth]{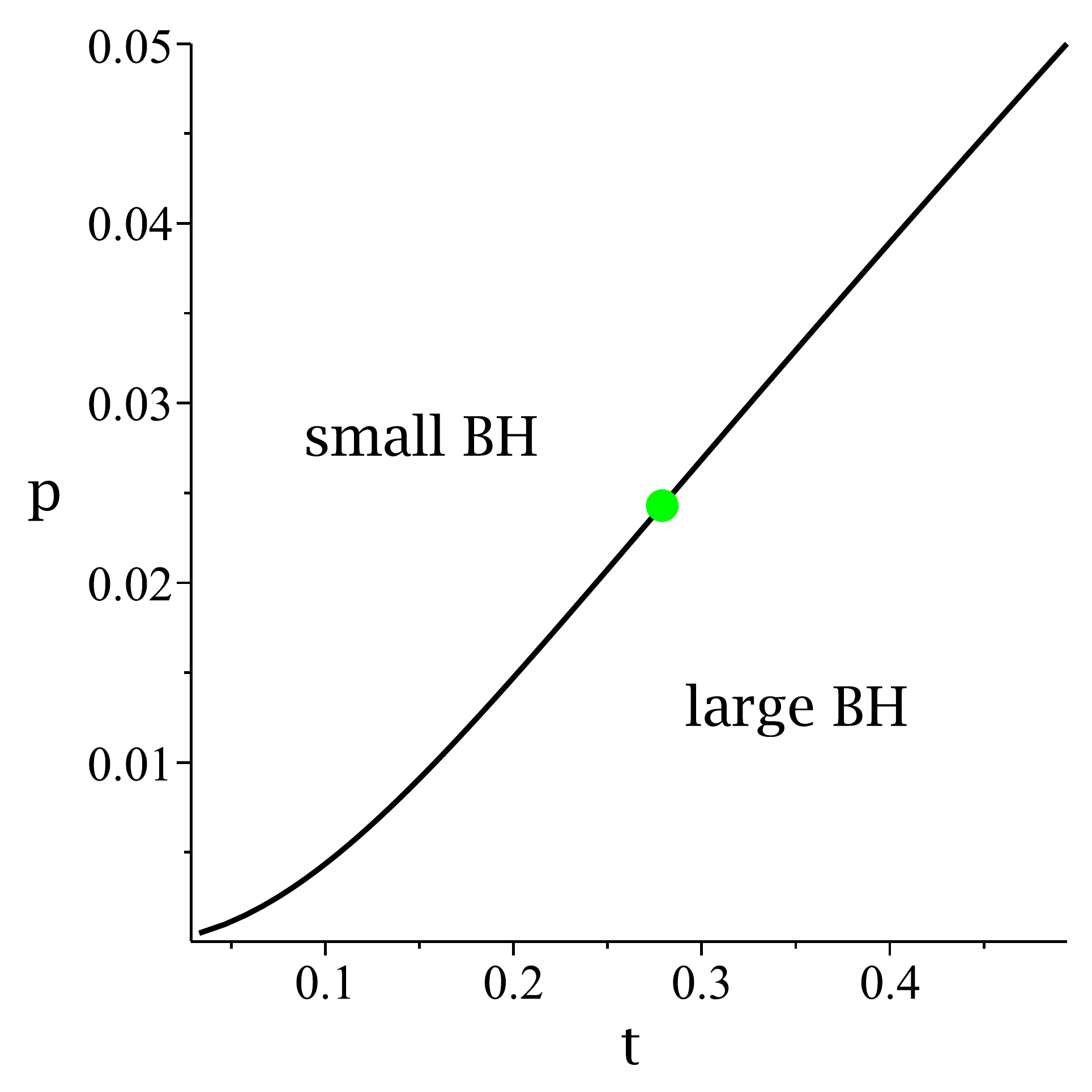}
\caption{{\bf Isolated critical point for spherical black holes}:  {\it Left}: $p$ vs. $t$ plot for $k=+1$ with $\thh \approx 1.002 \, \thh_{\delta=1,\epsilon=1}$.  {\it Right}: $p$ vs. $t$ plot for $k=+1$ with $\thh = \thh_{\delta=1, \epsilon=1}$.  As $\thh$ is adjusted to the value $\thh_{\delta=1, \epsilon=1}$ the two critical points (red dots) move together, merging for this particular value of $\thh$ (green dot). In both plots, $\alpha=-5$ and $\tq = 4$. }
\label{fig:icp_spherical}
\end{figure}

\begin{figure}[htp]
\centering 
\includegraphics[width=0.4\textwidth]{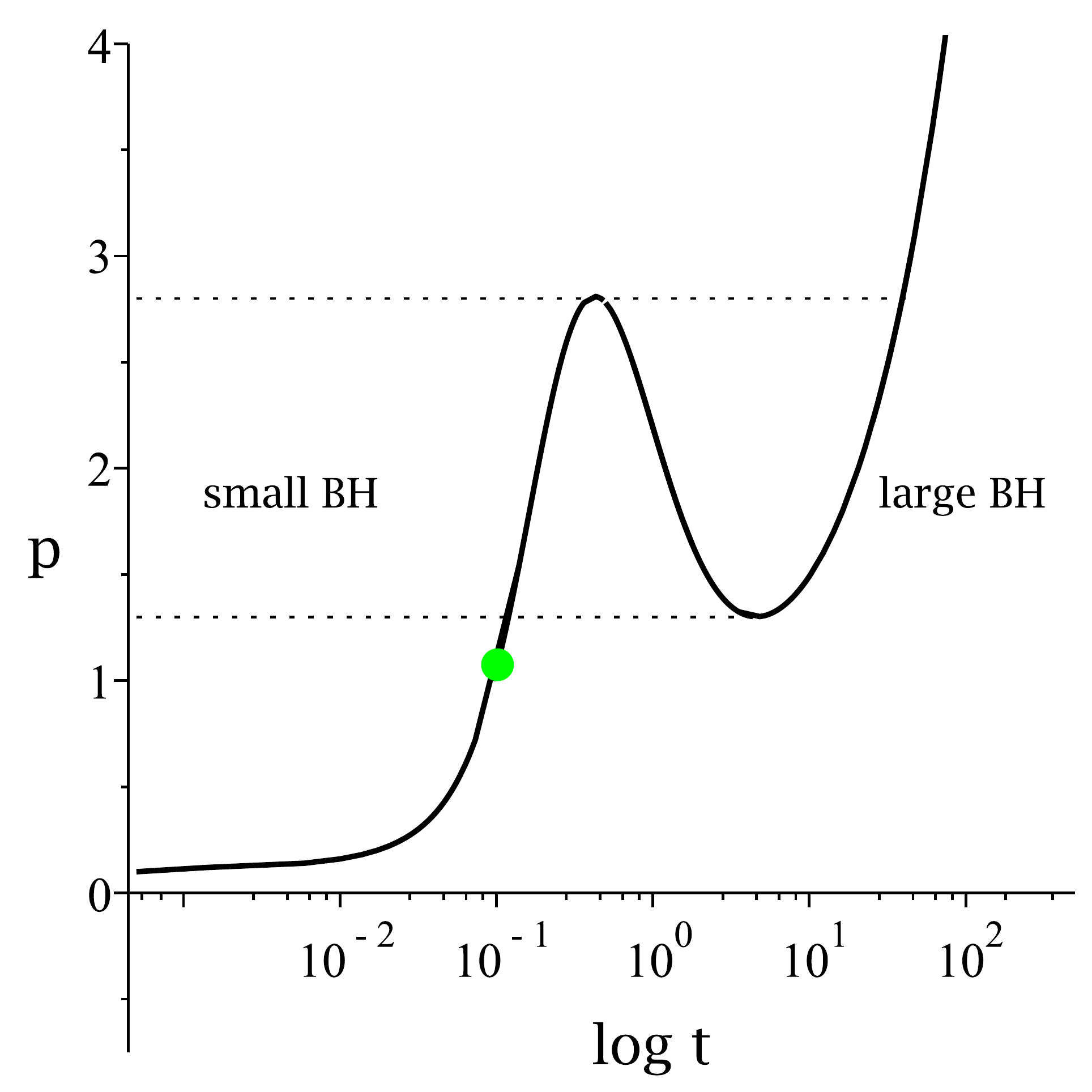}
\caption{{\bf Isolated critical point with double reentrant phase transition}:  Shown here for $q=0$ and $\alpha = -5$ with $\epsilon=-1$, the isolated critical point occurs for $\thh = \thh_{\delta=-1, \epsilon=-1} \approx 0.882797$ an isolated critical point (green dot) occurs along with a double reentrant phase transition.  The solid black line, which denotes a a coexistence curve for a first order phase transition, separates small and large black holes. For pressures between the two dotted lines, a small/large/small/large black hole phase transition occurs.}
\label{fig:icp_hyper}
\end{figure}

The phase diagrams for the isolated critical point in the spherical scenario are typical of what has been observed for this situation in previous work, and are portrayed in figure~\ref{fig:icp_spherical}.  In this left figure we see two separate critical points, each characterized by mean field theory critical exponents.  In the right plot, as the hair parameter is adjusted to $\thh = \thh_{\delta,\epsilon}$, the two critical points come together, ultimately coalescing to form the isolated critical point characterized by the non-standard critical exponents~\eqref{eqn:cl_critexponent1}.  

 In the hyperbolic case, even more interesting situations can occur.  For example, as shown in figure~\ref{fig:icp_hyper} for $\alpha = -5$, phase diagrams exist where an isolated critical point occurs along with a double reentrant phase transition.  In this plot, for the pressures between the two dotted lines, a small/large/small/large black hole reentrant phase transition occurs as the temperature is monotonically increased from zero.  This is the first example of this joint behaviour found to date.

\section{Coupling to the quasi-topological density}
\label{sec:qt_coupling}

Here we demonstrate that it is possible to conformally couple a scalar field to the quasi-topological density.  This allows for the coupling of the scalar field to  non-trivial cubic curvature terms  in five dimensions.  For scalar field configurations that are constant, the restrictions of couplings for spherical black holes lead to ghosty vacua \cite{EricksonRobie} unless the constant is zero;  coupling to cubic Lovelock terms can alleviate this but only in seven or more dimensions. We shall see that by coupling to the quasi-topological density, this issue can be alleviated in five dimensions as well. 

We apply the recipe of Oliva and Ray to the quasi-topological density, replacing the Riemann tensor and its contractions with $S_{\mu\nu\rho\lambda}$ and its contractions.  Thus, we consider the action 
\ba 
{\cal I} &=& \frac{1}{16\pi} \int d^5x \sqrt{-g} \bigg[-2\Lambda + R + \frac{\lambda}{2} {\cal L}_2 - \frac{7}{4} \mu{\cal Z}_5 - F^2
\nn\\ && \hspace{80pt}+ 16 \pi \left( b_0 \phi^5 S^{(0)} + b_1 \phi S^{(1)} + b_2 \phi^{-3} S^{(2)} + \mathfrak{b}_3 \phi^{-7} {\cal S}^{\rm QT} \right)  \bigg]
\ea 
where
\begin{align}
\mathcal{S}^{
\rm QT} &=  \left. \vphantom{\left( \frac{3(3D - 8)}{8} S_{\mu \alpha \nu \beta} S^{\mu \alpha \nu \beta} S \right.} {{{S_\mu}^\nu}_\alpha}^\beta {{{S_\nu}^\tau}_\beta}^\sigma {{{S_\tau}^\mu}_\sigma}^\alpha \right.  
               + \frac{1}{(2D - 3)(D - 4)} \left( \frac{3(3D - 8)}{8} S_{\mu \alpha \nu \beta} S^{\mu \alpha \nu \beta} S \right. \nonumber \\
              &  - 3(D-2) S_{\mu \alpha \nu \beta} {S^{\mu \alpha \nu}}_\tau S^{\beta \tau} + 3D\cdot S_{\mu \alpha \nu \beta} S^{\mu \nu} S^{\alpha \beta} 
               \left. + 6(D-2) {S_\mu}^\alpha {S_\alpha}^\nu {S_\nu}^\mu \right. \nonumber \\
              & \left. \left.- \frac{3(3D-4)}{2} {S_\mu}^\alpha {S_\alpha}^\mu S + \frac{3D}{8} S^3 \right) \right. \, .
\label{quasitop-hair}
\end{align}
Varying the action in $D=5$ with respect to the scalar field and using the ansatz,
\be\label{eqn:scalar_field_ansatz} 
\phi = \frac{K}{r}
\ee
results in the following equations  
\begin{align}
-24 \mathfrak{b}_3 k^3 +126 b_1 k K^4+35 b_0 K^6 &=0  
\nn\\
-12 \mathfrak{b}_3 k^2 +28 b_2 K^2 k +7 b_1 K^4 &=0 
\end{align}
which  not only  determine $K$, but also place a constraint  on the couplings, consistent with the situation considered earlier.

Evaluated on a general spherically symmetric metric, the field equations take the same form as eq.~\eqref{field-eqs},
\begin{align}\label{field-eqs} 
\left(1 - 2\frac{\lambda}{l^2} \kappa + \frac{3 \mu}{l^4} \kappa^2 \right)N'(r) &= 0
\nn\\
\left[3r^4\left(1 - \kappa + \frac{\lambda}{l^2} \kappa^2 - \frac{\mu}{l^4}\kappa^3 \right) \right]' 
&= \frac{2 e^2 r^3}{N^2(r)}\left[(rE)' \right]^2 - \frac{16 \pi l^2 h}{r^2} 
\nn\\
\left( \frac{r^3 (rE)'}{N(r)} \right)' &= 0
\end{align}
but now with
\be\label{eqn:qt_expr_h} 
h = \frac{b_0 K^6 + 6 b_1 k K^4 + \frac{24}{7} \mathfrak{b}_3 k^3}{K} \, .
\ee

 The calculation of the thermodynamic quantities can be repeated in a manner analogous to that done in section~\ref{sec:exact_soln}.  The form of the resulting expressions is identical to those given earlier with eq.~\eqref{eqn:qt_expr_h} substituted for $h$.  The only exception is the entropy, for which we find
\be 
S = \frac{A}{4} \left[1 + \frac{6k \lambda}{r_+^2} - \frac{9 k^2 \mu}{r_+^4} + 
\frac{16 \pi b_1 K^3}{r_+^3} + 
\frac{192 \pi k b_2 K}{r_+^3} + \frac{576 \pi k^2 \mathfrak{b}_3}{7 K r_+^3}   \right]
\ee
where
\be 
A = \omega_{3(k)} r_+^3 
\ee
is the area of the horizon.  Note that the hairy contribution to the entropy no longer permits a simple expression in terms of $h$.  Despite this, the presence of the $\mathfrak{b}_3$ terms will bring no novel thermodynamic behaviour into play, since the overall effect of the hair here is still simply the addition of a constant to the entropy.

It is also possible to couple to the quasi-topological density in higher dimensions, and we shall discuss this in the appendix.

\section{Linearized equations of motion and ghosts}
\label{sec:linearized_eqns}

Here we report on the linearized equations of motion of this theory where the real scalar field is coupled to terms up to the quasi-topological density.  Here we work keeping the dimension general.  Explicitly, we are working with the action,
\ba 
{\cal I} &=& \frac{1}{16\pi} \int d^D x \sqrt{-g} \bigg[-2\Lambda + R + \frac{\lambda}{(D-3)(D-4)} {\cal L}_2 + \frac{8(2D-3)}{(3D^2 - 15D + 16)(D-3)(D-6)}  \mu{\cal Z}_D 
\nn\\ &&  \hspace{30pt} - F^2 + 16 \pi \left( b_0 \phi^D S^{(0)} + b_1 \phi^{D-4} S^{(1)} + b_2 \phi^{D-8} S^{(2)} + \mathfrak{b}_3 \phi^{D-12} {\cal S}^{\rm QT}\right)  \bigg] \, , 
\ea
where the various terms in the Lagrangian are given by the expressions presented in sections~\ref{sec:exact_soln} and~\ref{sec:qt_coupling}.  

We consider perturbations about a maximally symmetric AdS background with Riemann tensor given by,
\be 
R_{abcd} = - 2 \frac{F_\infty}{l^2} \, g_{a[c} g_{b]d}
\ee
where $l$ sets the curvature scale of the AdS space and $F_\infty$ is the asymptotic value of the metric function.  This factor can be solved for by evaluating the field equations on this background,
\be 
1 - F_\infty + \frac{\lambda}{l^2} F_\infty^2 - \frac{\mu}{l^4} F_\infty^3 = 0 \, .
\ee

We take the scalar field to be a constant (possibly zero) in this background.  The equations of motion reduce to the following constraint for the constant scalar field $\Phi$:
\begin{align}
\label{eqn:constant_scalar} 
0 =& D b_0 \Phi^{D-1} - \frac{D(D-1)(D-2) b_1 F_\infty \Phi^{D-3}}{l^2} + \frac{D(D-1)(D-2)(D-3)(D-4) b_2 F_\infty^2 \Phi^{D-5}}{l^4}
\nn\\
&- \frac{D(D-1)(D-2)(D-3)(D-6)(3D^2 - 15D + 16) F_\infty^3 \mathfrak{b}_3\Phi^{D-7}}{8(2D-3) l^6} \, ,
\end{align}
while the linearized gravitational field equations take the form,
\begin{align} 
{\cal E}_{ab}^L &= \left[1 - 2 F_\infty \frac{\lambda}{l^2} + 3 F_\infty^2 \frac{\mu}{l^4} + 16 \pi b_1 \Phi^{D-2} - \frac{32 \pi (D-3)(D-4) F_\infty b_2 \Phi^{D-4}}{l^2} \right.
\nn\\
&+ \left. \frac{6 \pi (D-3)(D-6)(3D^2 - 15D + 16) F_\infty^2 \mathfrak{b}_3 \Phi^{D-6}}{(2D-3)l^4} \right]G^L_{ab}
\end{align}
where
\begin{align}
\label{eqn:linearized_feq} 
G^L_{ab} =& -\frac{1}{2} \bigg[ \square h_{ab}   + \nabla_{b}\nabla_{a}h^{c}{}_{c} -  \nabla_{c}\nabla_{a}h_{b}{}^{c} -  \nabla_{c}\nabla_{b}h_{a}{}^{c} 
  + g^{[0]}_{ab} \left(  \nabla_{d}\nabla_{c}h^{cd} -   \square h^{c}{}_{c} \right) 
  \nn\\
  &+ \frac{(d-1)F_\infty}{l^2} g_{ab}^{[0]}h^{c}{}_{c}  - \frac{2(D-1) F_\infty}{l^2} h_{ab}\bigg]
\end{align}
is the Einstein tensor (with cosmological term) linearized on this background.  Thus, unsurprisingly, the theory propagates the same degrees of freedom as Einstein gravity. The only additional constraint that must be imposed is the absence of ghosts, which demands the prefactor of $G_{ab}^L$ in eq.~\eqref{eqn:linearized_feq} must be positive.  The constraints reduce to the ordinary constraints for quasi-topological gravity in the case where the constant scalar field vanishes.  Considering eq.~\eqref{eqn:constant_scalar} we see that this is always possible provided either $ D > 7$ or $\mathfrak{b}_3$ vanishes.

We close with some remarks regarding more general non-zero scalar field backgrounds. In practice, the couplings $b_i$ are further constrained by demanding the existence of black hole solutions (see sections~\ref{sec:exact_soln} and~\ref{sec:qt_coupling} and the appendix).  In the case where the theory is expanded about a vanishing scalar field background, no additional constraints are added beyond what is already required by quasi-topological gravity.  In general, the theory admits pure AdS solutions with a constant scalar field.  However, we note that in the case where $\mathfrak{b}_3$ is absent, the existence of spherical black holes ($k=+1$) precludes the expansion about any background besides that in which the scalar field vanishes.  The reason is identical to that presented in~\cite{EricksonRobie}: the existence of spherical black holes requires $b_0b_1 < 0$, while the solution of eq.~\eqref{eqn:constant_scalar} (with $\mathfrak{b}_3=0$) requires $b_0b_1 > 0$.  Thus, when $\mathfrak{b}_3 = 0$ the background with vanishing scalar field is the unique configuration for which ghost free solutions can be constructed.  No such issues arise for the hyperbolic black holes, since there $b_0b_1 > 0$ suffices for both conditions.  

\begin{figure}[H]
\centering
\includegraphics[width=0.4\textwidth]{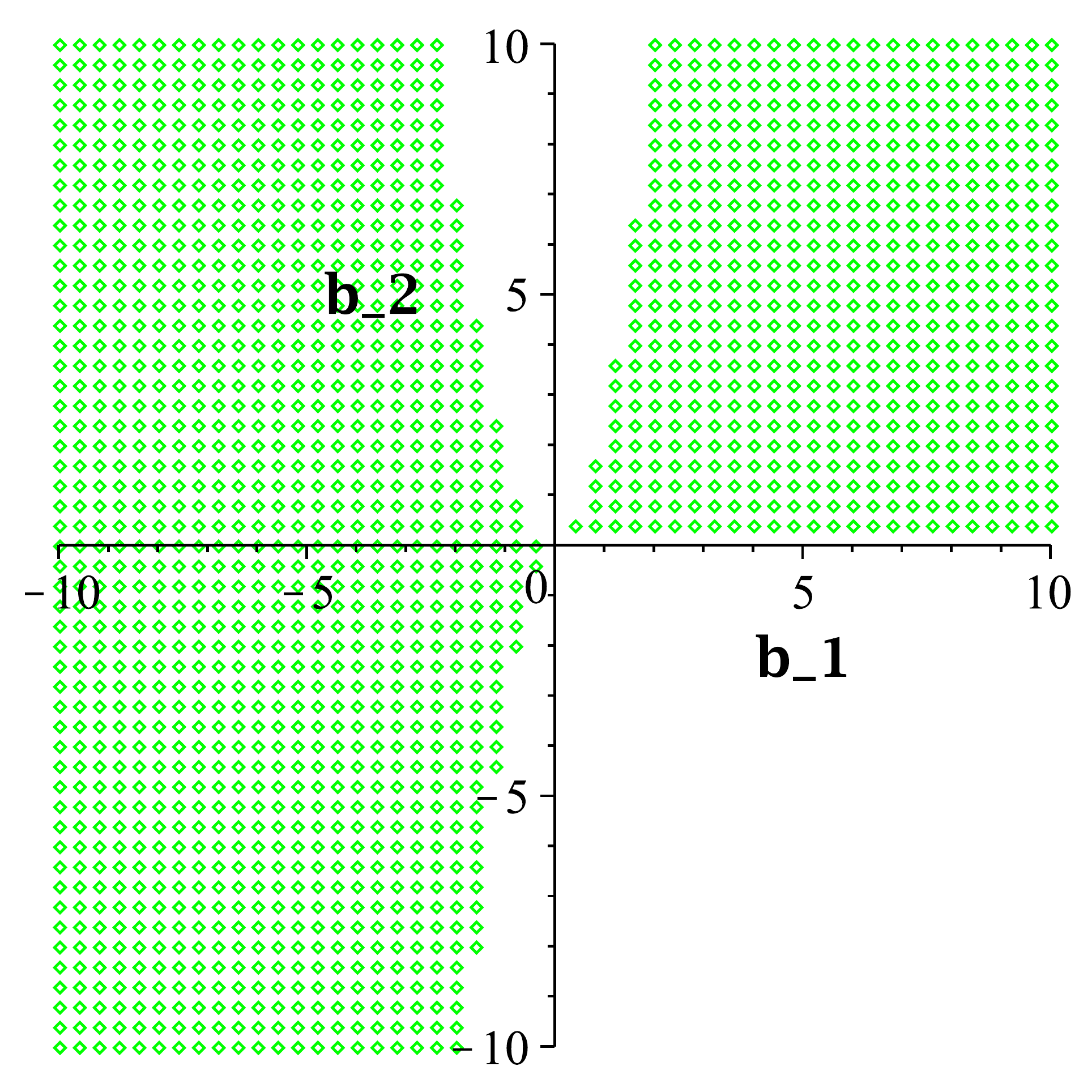}
\caption{{\bf Ghost-free parameter space}: An illustrative plot of ghost-free parameters for $D=5$ for the Einstein branch with $\tilde{b}_0 = 0.5, \tilde{\lambda} = 0.2$ and $\tilde{\mu} = 0.05$ for a non-zero, constant scalar field background.  The parameters indicated by green dots in the above plot correspond to ghost-free AdS vacua with non-zero constant scalar field under the constraints imposed on the couplings by the existence of spherical black holes [see eq.~\eqref{eqn:scalar_field_constraints}].}
\label{fig:ghost_free_parameters}
\end{figure}

When $\mathfrak{b}_3$ is non-zero, generic constant scalar field backgrounds are permitted in $D = 5$ or $ D \ge 7$ for both spherical and hyperbolic black holes.  In fact, when coupling the scalar field to the quasi-topological density, one can no longer consider the case of vanishing scalar field if $D < 7$, since in these cases the vanishing scalar field does not solve the equations of motion eq.~\eqref{eqn:constant_scalar}.  It is not hard to see, though, that configurations exist with AdS vacua with constant scalar field in which the effective Newton constant has the correct sign (i.e. the vacuum is ghost free).  However, since the equations involved are relatively complex, here we will simply demonstrate the existence of these solutions numerically for $D = 5$ and for the Einstein branch.  To this end, we work with dimensionless couplings, indicated by a tilde and defined by $\tilde{\lambda} = \lambda / l^2$, etc.  We show in figure~\ref{fig:ghost_free_parameters} an example ghost-free parameter space under the constraints imposed on the couplings by the existence of spherical black holes [see eq.~\eqref{eqn:scalar_field_constraints}].

\section{Conclusions}

We have conducted a systematic study of (charged) black holes in cubic quasi-topological gravity with a scalar field conformally coupled to the dimensionally extended Euler densities and the quasi-topological term. We have focused primarily on classifying the thermodynamics of these black holes in the framework of black hole chemistry, finding numerous examples of van der Waals behaviour, triple points and some novel phase structure including (multiple) reentrant phase transitions.

We have found that the black holes of this theory admit a black hole `$\lambda$-line': a line of second order phase transitions resembling that which marks the fluid/superfluid transition in liquid $^4$He. This phenomena was first observed for hairy black holes in Lovelock gravity in ref.~\cite{Hennigar:2016xwd} and the quasi-topological black holes meet the necessary conditions first presented in~\cite{Hennigar:2016xwd}.  As such, these black holes provide the second example of this phenomenon in black hole thermodynamics.  Finding further examples of black holes that can undergo this phase transition remains an interesting subject for further investigation.

We have also found the first examples of isolated critical points that occur for spherical black holes.  These critical points characterized by non-mean field theory critical exponents have so far been only observed to occur for black holes of hyperbolic horizon topologies.  Here, the quasi-topological black holes with scalar hair possess these critical points in both the spherical and hyperbolic cases.  In the hyperbolic cases, we have found that isolated critical points can occur simultaneously with multiple reentrant phase transitions.

Finally, we have investigated the theory when the scalar field is coupled to the cubic quasi-topological density as well as the dimensionally extended Euler densities. While this additional coupling does not affect the qualitative critical behaviour of the black holes, it does make contributions to the Wald entropy.  Furthermore, we find that this coupling allows for the existence of ghost-free AdS vacua in which the scalar field takes on a constant value---something that is not possible in the Lovelock case in five dimensions~\cite{EricksonRobie}.  

There remain a number of interesting directions for further study.  For example, it should be possible to construct general isolated critical points of the type first presented in~\cite{Dolan:2014vba} for five-dimensional, spherical black holes by considering higher order quasi-topological gravity~\cite{Dehghani:2013ldu, Cisterna:2017umf} conformally coupled to a scalar field.  Furthermore, the same considerations may lead to further examples of superfluid black hole behaviour and possibly to even more interesting results such as intersecting $\lambda$-lines.

\section*{Acknowledgments}

This work was supported in part by the Natural Sciences and Engineering Research Council of Canada.  

\appendix
\section{Hairy quasi-topological black holes in all dimensions}

Here we extend the $5$-dimensional solutions discussed in this work to all dimensions.  We couple the scalar field to up to the quasi-topological term, generalizing the solutions presented in section~\ref{sec:qt_coupling}.  Working with the action,
\ba 
{\cal I} &=& \frac{1}{16\pi} \int d^D x \sqrt{-g} \bigg[-2\Lambda + R + \frac{\lambda}{(D-3)(D-4)} {\cal L}_2 + \frac{8(2D-3)}{(3D^2 - 15D + 16)(D-3)(D-6)}  \mu{\cal Z}_D 
\nn\\ &&  \hspace{30pt} - F^2 + 16 \pi \left( b_0 \phi^D S^{(0)} + b_1 \phi^{D-4} S^{(1)} + b_2 \phi^{D-8} S^{(2)} + \mathfrak{b}_3 \phi^{D-12} {\cal S}^{\rm QT}\right)  \bigg] \, , 
\ea
and using the same conventions and notion as used earlier, we find that for a spherically symmetric ansatz the field equations take the form,
\begin{align}
\left[-1 + \frac{2\lambda}{l^2}\kappa - \frac{3 \mu}{l^4} \kappa^2 \right] N' &= 0 \, ,
\\
\left[(D-2)r^{D-1} \left(1 - \kappa + \frac{\lambda}{l^2} \kappa^2 - \frac{\mu}{l^4}\kappa^3 \right) \right]' & = \frac{2 e^2 r^{D-2}}{N^2} \left[(rE)' \right]^2 - \frac{16 \pi l^2 h}{r^2} \, , 
\\
\left[\frac{r^{D-2}(rE)'}{N} \right]' &= 0 \, .
\end{align}
where
\begin{align}
h &= b_0 K^D + (D-2)(D-3)b_1 k K^{D-2} + (D-2)(D-3)(D-4)(D-5) b_2 k^2 K^{D-4} 
\nn\\
&+ \frac{(D-2)(D-3)(D-6)(D-7)(3D^2-15D+16) \mathfrak{b}_3 k^3 K^{D-6}}{8(2D-3)}\, ,
\end{align}
and 
\be 
\phi = \frac{K}{r}\, .
\ee
These equations can be integrated easily -- taking $N = 1$ as a solution to the first equation, we obtain 
\be 
A = \sqrt{\frac{D-2}{2(D-3)}} \frac{e}{r^{D-3}} dt
\ee
for the electromagnetic one form and
\be 
1 - \tilde{\kappa} + \frac{\lambda}{l^2} \tilde{\kappa}^2 - \frac{\mu}{l^4} \tilde{\kappa}^3 = \frac{l^2 m}{r^{D-1}} - \frac{e^2 l^2}{r^{2(D-2)}} + \frac{16 \pi l^2 h}{(D-2) r^D} 
\ee
as the polynomial equation for the metric function. The equations of motion of the scalar field reduce to the following polynomial equations which must be satisfied
\begin{align}
0 &= K^6 D(D-1) b_0 + \frac{K^4(D-1)!(D(D-1)+4)k b_1}{(D-3)!} + \frac{K^2 (D-1)!(D(D-1) + 16) k^2 b_2}{(D-5)!} 
\nn\\
& + \frac{(D-1)(D-2)(D-3)(D-6)(D^2-D+36)(3D^2 - 15D + 16)k^3 \mathfrak{b}_3}{8(2D-3)}
\nn\\
0 &= \frac{K^4 (D-1)! b_1}{(D-3)!} + \frac{2K^2 (D-1)! k b_2}{(D-5)!} 
\nn\\
&+ \frac{3}{8} \frac{(D-1)(D-2)(D-3)(D-6)(3D^2-15D+16) k^2 \mathfrak{b}_3}{2D-3}
\end{align} 

Couplings to additional Lovelock terms is straightforward: the equations of motion of the scalar field are modified by the addition of extra terms from eq.~(2.17) of ref.~\cite{EricksonRobie}, and the hairy contribution to the field equations, $h$, becomes modified with the appropriate extra terms from eq.~(2.20) of ref.~\cite{EricksonRobie}.  We note that, in higher dimensions, the equation of state reduces to that for Lovelock gravity, thus no novel critical behaviour should be expected.

\bibliography{LBIB}
\bibliographystyle{JHEP}

\end{document}